\documentclass[a4paper,USenglish,cleveref, autoref, thm-restate,runningheads]{llncs}

\newcommand{\longversion}[1]{#1}
\newcommand{\shortversion}[1]{}
 
\usepackage[dvipsnames]{xcolor}
\usepackage{todonotes}
\usepackage{tikz}
\usetikzlibrary{arrows,shapes,snakes,automata,backgrounds,petri}
\usetikzlibrary{shapes.multipart}
\usepackage{comment}
\usepackage[noend]{algpseudocode}
\usepackage{caption}
\usepackage{subcaption}
\usepackage{hyperref}

\usepackage[utf8]{inputenc}          
\usepackage[T1]{fontenc}             
\usepackage[USenglish]{babel}        
\usepackage[intlimits]{amsmath}      
\usepackage{amsfonts}                
\usepackage{amssymb}                 
\usepackage{amsmath}

\shortversion{\usepackage{apxproof}[appendix=strip]
} 
\longversion{\usepackage[appendix=inline]{apxproof}}

\usepackage{mathtools}
\usepackage{setspace}
\usepackage{graphicx}
\usepackage{hyperref}
\usepackage{dsfont}
\usepackage{xspace}

\usepackage{algorithm}
\usepackage[noend]{algpseudocode}

\usepackage{algorithm}
\usepackage[noend]{algpseudocode}

\usepackage{lineno}
\longversion{\nolinenumbers} 
\shortversion{\nolinenumbers}

\newcommand{\EDS}{\textsc{Extension Dominating Set}\xspace}

\newcommand{\ERD}{\longversion{\textsc{Extension Roman Domination}}\shortversion{\textsc{Ext RD}}\xspace}

\newcommand{\ETRD}{\longversion{\textsc{Extension Total Roman Domination}}\shortversion{\textsc{Ext TRD}}\xspace}
\newcommand{\ECRD}{\longversion{\textsc{Extension Connected Roman Domination}}\shortversion{\textsc{Ext CRD}}\xspace}

\newcommand{\NP}{\textsf{NP}\xspace}
\newcommand{\Ptime}{\textsf{P}\xspace}
\newcommand{\FPT}{\textsf{FPT}\xspace}

\newcommand{\W}[1]{\ensuremath{\textsf{W}[#1]}}
\newcommand{\paraNP}{\textsf{para-NP}\xspace}

\newcommand{\nrdp}{\longversion{nice Roman domination property}\shortversion{nrdp}\xspace}
\newcommand{\nrdps}{\longversion{nice Roman domination properties}\shortversion{nrdps}\xspace}
\newcommand{\rdf}{\longversion{Roman dominating function}\shortversion{Rdf}\xspace}
\newcommand{\rdfs}{\longversion{Roman dominating functions}\shortversion{Rdfs}\xspace}
\newcommand{\EMRD}{\longversion{\textsc{Extension Maximum Roman Domination}}\shortversion{\textsc{Ext Max RD}}\xspace}

\newcommand{\hs}{\longversion{hitting set}\shortversion{hs}\xspace}

\newcommand{\ds}{\longversion{dominating set}\shortversion{ds}\xspace}

\newcommand{\mrdf}{\longversion{maximal Roman dominating function}\shortversion{mRdf}\xspace}
\newcommand{\mrdfs}{\longversion{maximal Roman dominating functions}\shortversion{mRdfs}\xspace}

\newcommand{\prdf}{\longversion{perfect Roman dominating function}\shortversion{pRdf}\xspace}

\newcommand{\crdf}{\longversion{connected Roman dominating function}\shortversion{cRdf}\xspace}
\newcommand{\crdfs}{\longversion{connected Roman dominating functions}\shortversion{cRdfs}\xspace}

\newcommand{\trdf}{\longversion{total Roman dominating function}\shortversion{tRdf}\xspace}
\newcommand{\trdfs}{\longversion{total Roman dominating functions}\shortversion{tRdfs}\xspace}

\newcommand{\RomanUpperbound}{1.9332}

\newcommand{\Oh}{\mathcal{O}}

\newcommand{\iffl}{if\longversion{ and only i}f\xspace}
\newcommand{\swlog}{\longversion{without loss of generality}\shortversion{w.l.o.g}\xspace}
\newcommand{\Wlog}{\longversion{Without loss of generality}\shortversion{W.l.o.g.}\xspace}

\makeatletter

\newenvironment{pf}{\begin{proof}}{\hfill\qed\end{proof}}
\newenvironment{pfclaim}{
\noindent\emph{Proof.}}
{\hfill$\Diamond$}

\begin{document}
\title{Enumeration With Nice Roman Domination Properties}
\author{Kevin Mann
\orcidID{0000-0002-0880-2513} 
}
\authorrunning{Kevin Mann}
\institute{
Universit\"at Trier, Fachbereich~4 -- Abteilung Informatikwissenschaften\\  
54286 Trier, Germany.\\
\email{mann@uni-trier.de}
}

\maketitle\begin{abstract}
Although  \textsc{Extension Perfect Roman Domination} is \NP-complete, all minimal (with respect to the pointwise order) perfect Roman dominating functions can be enumerated with polynomial delay. This algorithm uses a bijection between minimal perfect Roman dominating functions and Roman dominating functions and the fact that all minimal Roman dominating functions can be enumerated with polynomial delay. This bijection considers the set of vertices with value~2 under the functions.
In this paper, we will generalize this idea by defining so called \emph{nice Roman Domination properties} for which we can employ this method. With this idea, we can show that all minimal \emph{maximal Roman Dominating functions} can be enumerated with polynomial delay in $\mathcal{O}(\RomanUpperbound^n)$ time. Furthermore, we prove that enumerating all minimal connected/total Roman dominating functions on cobipartite graphs can be achieved with polynomial delay. Additionally, we show the existence of a polynomial-delay algorithm for enumerating all minimal connected Roman dominating function on interval graphs. We show some downsides to this method as well.
\end{abstract}

\section{Introduction}

Enumeration problems have many different applications, as for example data mining, machine learning \cite{GunKMT97} or biology/chemistry~\cite{Mar2015}. One possible property of enumeration algorithms is polynomial delay. 
An enumeration algorithm has polynomial delay if the time between two new outputs is polynomially bounded. This is interesting for multi-agent systems, where one agent enumerates and the other processes the output. In this scenario, we know that the second agent does not need to wait for too long and use unnecessary resources. One example for polynomial delay enumeration is enumerating minimal \rdfs \cite{AbuFerMan2024}. This is interesting, since it is open for more than~4 decades if there exists an enumeration algorithm for minimal dominating sets with polynomial delay (or even output polynomial time). 

\textsc{Roman Domination} is a well-studied variation of \textsc{Dominating Set}, see \longversion{\cite{Ben2004,ChaCCKLP2013,Cocetal2004,Dre2000a,Fer08,Lie2007,Lieetal2008,LiuCha2013,PenTsa2007,ShaWanHu2010}}\shortversion{\cite{ChaCCKLP2013,Cocetal2004,Lieetal2008,LiuCha2013}}, being motivated by a defense strategy of the Roman Empire~\cite{Ste99}. In this scenario, you want to put up to~2 legions per region such that each region has at least one legion on itself or there is a neighbored region with two legions on it. This can be modeled on graphs where each region is represented by its own vertex and  vertices are connected  \iffl the corresponding regions are neighbors. The distribution of the legions can be represented by a function $f\colon V\to \{0,1,2\}$ for the set of regions~$V$. $f\colon V\to \{0,1,2\}$ is called a \emph{Roman dominating function} (Rdf for short) on a graph $G=(V,E)$ if, for all $v\in V$ with $f(v) =0$, there is a $u\in V$ with $f(u)=2$ and $\{v,u\} \in E$.
A function $f\colon V\to \{0,1,2\}$ can also be represented as a partition of $V$ into $(V_0(f),V_1(f),V_2(f))$, where $V_i(f)\coloneqq \{v\in V\mid f(v)=i\}$ for all $i\in \{0,1,2\}$.

The polynomial delay algorithm in \cite{AbuFerMan2024} uses \ERD, which is polynomial time solvable. Using the extension version of a problem is a usual strategy to obtain a polynomial-delay algorithm. In this paper, we will only consider monotonically increasing problems\longversion{ (We enumerate minimal elements with respect to an ordering)}. Therefore, we will explain the idea of extension problems with respect to these. For more information, we refer to~\cite{CasFGMS2022}. The idea of an extension problem is to find a minimal solution which is greater than or equal to a given presolution with respect to the given partial order. Then, \textsc{Extension X Roman Domination} (where X specifies the variant, which is empty in the classical version) is defined as follows:

\noindent
\centerline{\fbox{\begin{minipage}{.96\textwidth}
\textbf{Problem name: }\textsc{Extension X Roman Domination}, or  \textsc{Ext X RD} for short.\\
\textbf{Given: } A graph $G=(V,{E})$ and a function $f\in \{0,1,2\}^V$.\\
\textbf{Question: } Is there a minimal X \rdf $g$ on $G$ with $f \leq g$?
\end{minipage}}}

Here, $f\leq g$ is understood pointwisely: for all $v\in V$, $f(v)\leq g(v)$.
In \cite{ManFer2024a}, it is proven that enumerating minimal perfect Roman dominating functions (or pRdfs for short; a pRdf is a \rdf $f$ such that all vertices in $V_0(f)$ have exactly one neighbor in $V_2(f)$) can be done with polynomial delay even though the extension version is \NP-complete. The enumeration algorithm is based on a bijection $B$ between minimal \rdfs and minimal \prdf on the same graph. For a minimal \rdf $f$, $V_2(f)=V_2(B(f))$. In this paper, we want to generalize this idea for more Roman domination variations like total, connected or maximal.  
To this end, we will separate the choice of vertices with value~1 and value~2. Therefore, we define nice Roman domination properties in \autoref{sec:nrdp}. We also prove how to make use of this by enumerating all minimal functions of a variation with a nice Roman domination property and a fixed set of vertices with value~2. From this situation, some enumeration properties can be inherited to enumerating all minimal functions of this variation. 
In \autoref{sec:enum_mrdf}, we use these results to prove that all minimal \mrdfs can be enumerated with polynomial delay and in $\mathcal{O}(\RomanUpperbound^n)$. We use this technique also for enumerating minimal \trdfs/\crdfs on cobipartite in \autoref{sec:enum_tcrdf}. This section also includes a polynomial delay algorithm for enumerating minimal \crdfs in interval graphs. We also show that this approach sometimes fails unless $\Ptime=\NP$ in \autoref{sec:downsides_nrdp}. 
\shortversion{

Some proofs can be found in the appendix. The corresponding results are marked by $(\star)$.\todo{Arxiv}}

\section{Definitions and Notation}

By $\mathbb{N}=\{0,1,2,\ldots\}$ we denote the non-negative integers. For all $n\in \mathbb{N}$, $[n]\coloneqq \{1,\ldots,n\}$. $\binom{A}{k}$ denotes  all subsets of a set $A$ of size exactly $k\in \mathbb{N}$.  For two sets $A,B$, the set of all functions $f:A\to B$ is denoted by $B^A$. A special function is the \emph{characteristic function} $\chi_A:B\to \mathbb{N}$ such that for all $v \in B$, $\chi_A (v)=1$ \iffl $v\in A$ and $\chi_A(v)=0$ otherwise. 
 
In this paper, we only consider simple undirected graphs $G=(V,E)$, where $V$ denotes the vertex set and $E\subseteq \{\{v,u\}\mid v,u\in V\}$ is the set of edges. The by $A\subseteq V$ induced subgraph of $G$ is defined as $G[A]=(A,E \cap \binom{A}{2})$. For all $v\in V$, $N(v)\coloneqq \{u\mid \{v,u\}\in E\}$ is called the (open) neighborhood of $v$ with respect to~$G$. The closed neighborhood of $v\in V$ is denoted by $N[v]\coloneqq \{v\} \cup N(v)$. For a set $A\subseteq V$, we define the neighborhood by the union $N(A)=\bigcup_{v \in A}N(v)$ or $N[A]=\bigcup_{v \in A}N[v]$. $P_{G,A}(v)=N[v] \setminus N[A\setminus \{v\}]$ is the set of private neighbors of $v\in A\subseteq V$ with respect to $A$ on $G$. We also consider hypergraphs $H=(X,S)$, where $X$ is the\longversion{ so-called} universe and $S\subseteq 2^X$ is the set of hyperedges. 

We call a set $D\subseteq V$ a dominating set of the graph $G=(V,E)$ if, for all $ v\in V$, $N[v] \cap D\neq \emptyset$. This can be viewed as a special case of hitting sets. A hitting set $D\subseteq X$ of the hypergraph $H=(X,S)$ if for all $s\in S$, $s\cap D\neq \emptyset$. There are many different variations of dominating sets on a graph $G=(V,E)$, as connected dominating set, cds for short, ($D$ has to be connected), or total dominating set ($D$ is a hitting set of $(V,\{N(v)\mid v\in V\})$).
Similar to dominating set, Roman domination has also many variations\shortversion{. Some of these are mentioned in the appendix. In this paper, we will only consider maximal/total/connected \rdf, extensively.} 
\begin{toappendix} 
\shortversion{\subsection{Roman Domination Variations}
Here are some variations of \textsc{Roman Domination}:}

\begin{itemize}
    \item \textsc{Italian Domination} or \textsc{Roman $\{2\}$-domination} ($f\colon V\to \{0,1,2\}$ is an Italian dominating function if for each $v\in V_0(f)$, the values under~$f$ of the neighbors sum up to at least~2) \cite{CheHHM2016,HenKlo2017};
    \item \textsc{Perfect Roman Domination} ($f\colon V\to \{0,1,2\}$ is a perfect Roman dominating function if for each $v\in V_0(f)$, there exists exactly one neighbor of value~2)~\cite{HenKloMac2018};
    \item \textsc{Unique Response Roman Domination} ($f\colon V\to \{0,1,2\}$ is a unique response Roman dominating function if $f$ is a \prdf and for each $v\in V_1(f)\cup V_2(f)$, $v$ has no neighbor in $V_2(f)$)~\cite{RubSla2007};
    \item \textsc{Total Roman Domination} ($f\colon V\to \{0,1,2\}$ is a total Roman dominating function if $f$ is a \rdf and $G[V_1(f)\cup V_2(f)]$ has no isolates)~\cite{AbdHSY2016}
    \item \textsc{Connected Roman Domination} ($f\colon V\to \{0,1,2\}$ is a connected Roman dominating function if $f$ is a \rdf and $G[V_1(f)\cup V_2(f)]$ is connected)~\cite{LiRZD2022}.
    \item \textsc{Maximal Roman Domination} (the \rdf $f\colon V\to \{0,1,2\}$ is a maximal Roman dominating function if $V_0(f)$ is no dominating set)~\cite{ShaSLDJ2024}.
\end{itemize} 
\end{toappendix}

Further, we consider some notation of enumeration. As for decision problems, we need an alphabet~$\Sigma$. Let $A\subseteq \Sigma^* \times \Sigma^*$ be a binary relation. An enumeration problem \textsc{Enum $A$} is defined as enumerating all elements in $A(x) \coloneqq \{y\in \Sigma^* \mid A(x,y)\}$ for a given instance $x\in \Sigma^*$ without repetitions.

The analysis of the running time for enumeration is divided into input-sensitive (running time with respect to the input size $\vert x\vert$) and output-sensitive (running time with respect to the input and output size $\vert x \vert + \vert A(x) \vert$).
We mentioned already polynomial delay which is a output-sensitive property. For a polynomial delay algorithm there exists a polynomial $p\colon \mathbb{N} \to \mathbb{N}$ such that the running time to the first output, between two consecutive outputs and after the last is bounded in $p(\vert x\vert)$ for the input~$x$. Another possible property of an enumeration algorithm is output-polynomial time where there exists a polynomial $p\colon \mathbb{N} \times \mathbb{N} \to \mathbb{N}$ such that the running time is bounded in $p(\vert x\vert, \vert A(x)\vert)$ for the input~$x$. It should be mentioned that a polynomial delay algorithm is also output polynomial. Enumerating minimal hitting sets on a hypergraph, also known as minimal transversal enumeration problem (denoted by \textsc{Enum Tr}), is a very famous enumeration problem, for which it is open for more than four decades whether there is an output-polynomial time algorithm. 

As for classical complexity, there are also reductions for enumeration problems. \shortversion{For example \emph{parsimonious reduction}~\cite{Str2019} and \emph{e-reduction}~\cite{CreKPSV2019}. The definition to both can be found in the appendix.} 
\begin{toappendix}
\shortversion{\subsection{Enumeration Reductions}}
Now we consider two \longversion{of these.}\shortversion{reductions for enumeration problems.} For this purpose let \textsc{Enum $A$} and \textsc{Enum $B$} be two enumeration problems. A pair of polynomial-time computable  functions $f, g$, is called a \emph{parsimonious reduction} if, for each instance $x$ of \textsc{Enum $A$},  $f(x)$ is an instance of \textsc{Enum $B$} such that $g(x)$ is a bijection between $A(x)$  and $B(f(x))$.  See \cite{Str2019} for more details.

The idea of our technique will be based on the notion of an \emph{e-reduction}~\cite{CreKPSV2019}: Let $\Sigma$ be an alphabet, $\textsc{Enum A},\textsc{Enum B}\in \textsf{EnumP}$ be enumeration problems over the alphabet $\Sigma$. We say \textsc{Enum A} \emph{e-reduces} to \textsc{Enum B} ($\textsc{Enum A}\leq_e\textsc{Enum B}$ for short) if there exists a polynomial-time computable function $f:\Sigma^*\to \Sigma^*$ as well as a relation $\tau\subseteq \Sigma^* \times \Sigma^*\times \Sigma^*$, such that the following conditions hold:
\begin{enumerate}
    \item $A(x)= \bigcup_{y\in B(f(x))} \tau(x,y,\cdot)$, where $\tau(x,y,\cdot)\coloneqq\{z\in\Sigma^*\mid \tau(x,y,\cdot)\}$ for all $x,y\in \Sigma^*$,
    \item For all $y\in B(f(x))$, either $\tau(x,y,\cdot)=\emptyset$ or $\emptyset \subsetneq \tau(x,y,\cdot)\subseteq A(x)$ and $\tau(x,y,\cdot)$ can be enumerated with polynomial delay in~$\vert x \vert$; moreover $\vert \{ y\in B(f(x)) \mid \tau(x,y,\cdot)=\emptyset\}\vert$ is polynomial bounded in $\vert x \vert$.  
    \item For all $z\in A(x)$, $ \{y\in\Sigma^*\mid \tau(x,y,z)\}\subseteq B(f(x))$ and $\vert \tau(x,y,\cdot)\vert$ is bounded polynomially in~$\vert x \vert$.
\end{enumerate}
\end{toappendix}

\section{Nice Roman Domination Property}\label{sec:nrdp}

In this section, we will set the theoretical basis of this paper. The main goal is to prove \autoref{thm:nrdp_polyenumeration}. For this paper, we assume that a property $\mathcal{P}$ is a set of functions mapping the vertex set of a graph to $\{0,1,2\}$. Instead of $f\in \mathcal{P}$, we say $f$ has property $\mathcal{P}$.\footnote{Actually, we have to say $f$ has property $\mathcal{P}$ with respect to a graph $G$, but the graph is known due to the context}. To this end, we start by defining nice Roman domination properties.
We call a property $\mathcal{P}$ a \emph{nice Roman domination property} (\emph{nrdp} for short) if for every $G=(V,E)$ and every $f\in \{0,1,2\}^V$ with property $\mathcal{P}$ fulfills the following constraints:
\begin{enumerate}
    \item $f$ is a \rdf.\label{con:nrdp_rdf}
    \item For all $v\in V_0(f)$, $f+\chi_{\{v\}}$ has property $\mathcal{P}$.\label{con:nrdp_add_from_V_0}
    \item For all $v\in V_2(f)$, $f- \chi_{\{v\}}$ has property $\mathcal{P}$ \iffl $P_{G[V\setminus V_1(f)],V_2\left(f\right)}\left( v \right) \subseteq \lbrace v\rbrace$, i.e. $f-\chi_{\{v\}}$ does not have property $\mathcal{P}$ \iffl $v$ has a private neighbor, but itself, on the graph $G[V_0(f) \cup V_2(f)]$.\label{con:nrdp_V_2}
\end{enumerate}

For a \nrdp $\mathcal{P}$, a graph $G$ and an $A\subseteq V$, define $C_{\mathcal{P},G}[A]$ as the set of minimal $f\in \{0,1,2\}^V$ with respect to $\leq$ and the property~$\mathcal{P}$ such that $V_2(f)=A$. Now, we are ready to formulate our main theorem.

\begin{theorem}\label{thm:nrdp_polyenumeration}
    Let $\mathcal{P}$ be a \nrdp and $\mathcal{G}$ be a class of graphs. If there exists an algorithm~$\mathcal{A}$ that enumerates all elements in $C_{\mathcal{P},G}[A]$ in output-polynomial time (resp. with polynomial delay) for all $G=(V,E) \in \mathcal{G}$ and $A\subseteq V$, then there exists an algorithm to enumerate all minimal elements  with property $\mathcal{P}$ on any $G\in \mathcal{G}$ in output-polynomial time (resp. with polynomial delay).
\end{theorem} 

\noindent
\autoref{thm:nrdp_polyenumeration} is a modification of $\leq_e$-reductions\shortversion{~\cite{CreKPSV2019}} for \nrdp to get stronger results and even output-polynomial time. \shortversion{For the proof,  we first show a couple of lemmas.}
\longversion{In order to prove this theorem, we first show a couple of lemmas.} The next lemma follows from \cite[Lemma 7]{AbuFerMan2024}.

\begin{lemma}\label{lem:bij_min_rdf_set}
    Let $G=(V,E)$ be a graph. $F(A)\coloneqq2\cdot \chi_A + \chi_{V\setminus N[A]}$ is a bijection between all sets $A\subseteq V$ such that for all $v\in V$, $ P_{G[N[A]],A}(v)\nsubseteq \{v\}$ and all minimal \rdfs of $G$.
\end{lemma}

\begin{lemma}\label{lem:NRDP_G_subset}
    Let $\mathcal{P}$ be a \nrdp, $G=(V,E)$ be graph and $A\subseteq B\subseteq V$. If $C_{\mathcal{P},G}[B]\neq \emptyset$ then $C_{\mathcal{P},G}[A]\neq \emptyset$.
\end{lemma}

\begin{pf}
    Let $g\in C_{\mathcal{P},G}[B]$ and for each $v\in B$ let $Y_v= P_{G, B}(v)\cap V_0(g)$. Define $Y_A \coloneqq \bigcup_{v\in A}Y_v$ and $f\coloneqq  \chi_{V\setminus Y_A} + \chi_{A}$. We want to show that $f$ has property~$\mathcal{P}$. Observe $V_0(f)=Y_A \subseteq N(A)$. Hence, $f$ is a \rdf. By Constraint~\ref{con:nrdp_add_from_V_0} of \nrdp and adding $\chi_{\{u\}}$ for $u\in V_0(g) \setminus Y_A$ to $g$, $\widetilde{g} \coloneqq g + \chi_{V_0(g) \setminus Y_A}$ has property $\mathcal{P}$. Furthermore, $\widetilde{g}= f + \chi_{B\setminus A}\geq f$, as $V_2(\widetilde{g}) = B = V_2(f) \cup B\setminus A$, $V_0(\widetilde{g})=Y_A= V_0(f)$ and $B\setminus A \subseteq V_1(f)$ as they are no private neighbors of elements in $A$ with respect to $B$. By definition of~$f$ and~$Y_A$, $N(u)\cap V_0(f)=\emptyset$ for all $u \in B\setminus A$. By an inductive argument, Constraint \ref{con:nrdp_V_2} of \nrdp implies $\widetilde{g} - \chi_{B\setminus A} = f$ has property~$\mathcal{P}$.

    We now show that from~$f$, we can construct some $h\in C_{\mathcal{P},G}[A]$. If $f$ is minimal, set $h=f$. Otherwise, assume there is a $h\in \{0,1,2\}^V$ with property~$\mathcal{P}$ on~$G$, $h \leq f$ and a $u\in V$ such that $h(u)<f(u)$. If $f(u)=2$ then the elements in~$Y_u$ are not dominated by~$h$, as they private neighbors of $u$. This is not possible and, therefore, $V_2(h)=V_2(f)$. This implies $h=f-\chi_M$ for an $M\subseteq V_1(f)$, $M\neq\emptyset$, and $h\in C_{\mathcal{P},G}[A]$. Thus, $C_{\mathcal{P},G}[A] \neq \emptyset$.
\end{pf}

\noindent
Now we will show a connection between $C_{\mathcal{P},G}$ and extension problems.

\begin{lemma}\label{lem:NRDP_G_eq_extension}
    Let $G=(V,E)$ be a graph and $A\subseteq V$. $C_{\mathcal{P},G}[A]\neq \emptyset$ \iffl there is a minimal  $g \in \{0,1,2\}^V$ with property $\mathcal{P}$ on $G$ such that $2\cdot \chi_A + \chi_{V\setminus N[A]}\leq g$. Further, for all $h\in C_{\mathcal{P},G}[A]$, $2\cdot \chi_A + \chi_{V \setminus N[A]} \leq h$.
\end{lemma}

\begin{pf}
    Let $C_{\mathcal{P},G}[A]\neq \emptyset$. By \autoref{lem:bij_min_rdf_set}, $f\coloneqq 2\cdot \chi_A + \chi_{V\setminus N[A]}$ is the only minimal \rdf with $V_2(f)= A$. Thus, every \rdf $g$ on $G$ with $V_2(g)=A$ fulfills $f \leq g$. 
    Conversely, let $g\in \{0,1,2\}^V$ be minimal with the property $\mathcal{P}$ on $G$ with $f\leq g$. This implies $V_0(g)\subseteq V_0(f)= N[A] \setminus A$. If there exists an $x\in V_2(g) \setminus A$, $x$ has no private neighbor in $V_0(g)$ and $g$ is not minimal (see Constraint \ref{con:nrdp_V_2} of \nrdps). Hence, $C_{\mathcal{P},G}[A]\neq \emptyset$ and for all $g\in C_{\mathcal{P},G}[A]$, $f\leq g$.
\end{pf}

Let us shortly follow a corollary before proving \autoref{thm:nrdp_polyenumeration}. We use this result to discuss the why our approach is helpful.

\begin{corollary}\label{cor:Ext_nrdp_1_2_notneighbor}\shortversion{$(\star)$}
    Let $\mathcal{P}$ be a \nrdp, $G=(V,E)$ a graph and $f\in \{0,1,2\}^V$ with $N(V_1(f))\cap V_2(f)=\emptyset$. Then, there is an $h\in \{0,1,2\}$ minimal with the property $\mathcal{P}$ and $f\leq h$ \iffl $C_{\mathcal{P},G}[V_2(f)]\neq \emptyset$.
\end{corollary}

\begin{toappendix}
    \begin{pf}\shortversion{(to \autoref{cor:Ext_nrdp_1_2_notneighbor})}
    The if part follows by \autoref{lem:NRDP_G_eq_extension} and $f\leq 2\cdot \chi_A + \chi_{V \setminus N[A]}$. So, let us assume there is minimal $h\in \{0,1,2\}$ with property $\mathcal{P}$ with $f\leq h$ but $C_{\mathcal{P},G}[V_2(f)] = \emptyset$. Then $V_2(f)\subseteq V_2(h)$ and $C_{\mathcal{P},G}[V_2(h)]\neq \emptyset$, which contradicts \autoref{lem:NRDP_G_subset}.
\end{pf}
\end{toappendix}

\noindent
Now we can turn our attention back to \autoref{thm:nrdp_polyenumeration}.

\begin{pf}[of \autoref{thm:nrdp_polyenumeration}]
    We will only consider the output-polynomial time case. The proof for the polynomial-delay case is analogous. Hence, let $\mathcal{A}$ be the algorithm for enumerating $C_{\mathcal{P},G}[A]$ for given $G=(V,E)\in \mathcal{G}$ and $A\subseteq V$ and $p$ the polynomial such that $p(\vert V\vert +\vert E \vert ,\vert C_{\mathcal{P},G}[A]\vert)$ is the running time of the algorithm. \Wlog, all the multiplicative constants of the monomials are positive. Otherwise, we could use the polynomial with the absolute values of the multiplicative constants instead. 
    
    The idea of the algorithm for this proof is based on a branching algorithm. We\longversion{ will} branch on each vertex if it is in~$A$. Every time we add a vertex to~$A$, we compute~$C_{\mathcal{P},G}[A]$. If $C_{\mathcal{P},G}[A]=\emptyset$, we can skip this branch. This case can be checked in polynomial time, as we can enumerate all elements in output-polynomial time. If $C_{\mathcal{P},G}[A] \neq \emptyset$, we enumerate the elements and go on  with the next vertex. 

    Now we want to prove that this algorithm works. So, let $G=(V,E)\in \mathcal{G}$. Clearly, $\bigcup_{A\subseteq V}C_{\mathcal{P},G}[A]$ is exactly the set of all minimal elements with property~$\mathcal{P}$ on~$G$. By \autoref{lem:NRDP_G_subset}, we can skip the branches with $C_{\mathcal{P},G}[A] = \emptyset$. Therefore, we only need to show that the algorithm runs in output-polynomial time. To do this, we will prove the following claim:
    \begin{claim}\label{cla:NRDP_consecutive_empty}
        The algorithm  considers at most $\vert V\vert^2$ many consecutive sets $A \subseteq V$ with $C_{\mathcal{P},G}[A]= \emptyset$ before the algorithm either stops or finds a $B\subseteq V$ with $C_{\mathcal{P},G}[B] \neq \emptyset$.
    \end{claim}
    \begin{pfclaim}
        If $C_{\mathcal{P},G}[\emptyset]=\emptyset$ then the algorithm stops directly by \autoref{lem:NRDP_G_subset}. So, assume there is a $B' \subseteq V$ such that $C_{\mathcal{P},G}[B']$ is not empty. In this case, 
        the algorithm would now go through at most $\vert V\vert$ vertices which it did not consider so far to obtain $B'$. For each of these vertices $v\in V\setminus B'$, the algorithm either has found $C_{\mathcal{P},G}[B' \cup \{v\}]\neq \emptyset$ or would cut this branch and go on with the next vertex. 

        After this, the algorithm goes through all of the following cases: it deletes an element from $B'$ and checks as before. It will go on with this process until it either finds a set $B$ with $C_{\mathcal{P},G}[B]\neq \emptyset$ or it stops. As the algorithm can only delete at most $\vert B'\vert \leq \vert V\vert $ vertices from $B'$ and after each deletion step it has to consider at most $\vert V\vert $ before deleting the next vertex, the algorithm considers at most $\vert V\vert^2 $ consecutive sets $A\subseteq V$ with $C_{\mathcal{P},G}[A]= \emptyset$ before stopping.
    \end{pfclaim}
    
    Let $\mathcal{B}\coloneqq \{B\subseteq V \,\vert\, C_{P, G}[B]\neq \emptyset\}$.
    The claim implies that we consider at most $\vert V\vert ^2\cdot \vert \mathcal{B} \vert$ sets $A\subseteq V$ with $C_{\mathcal{P},G}[A]=\emptyset$. Hence, the running time is bounded~by 
    
    \begin{equation*}
        \begin{split}
            &\left(\sum_{A\in \mathcal{B}} p(\vert V\vert +\vert E\vert, \vert C_{\mathcal{P},G}[A] \vert)\right) + \vert V\vert ^2\cdot \vert  \mathcal{B} \vert \cdot p(\vert V\vert+\vert E\vert ,0) \\
            \leq & \left(\sum_{B\in \mathcal{B}} \vert C_{\mathcal{P},G}[B] \vert\right)\cdot \left( p(\vert V\vert +\vert E\vert, \sum_{B\in \mathcal{B}} \vert C_{\mathcal{P},G}[B] \vert) + \vert V\vert ^2 \right)\cdot p(\vert V\vert+\vert E\vert ,0).
        \end{split}
    \end{equation*}
    This is a polynomial in $\vert V\vert+ \vert E\vert $ and the output size.
\end{pf}

It is important for the polynomial delay case that in the claim of  \autoref{thm:nrdp_polyenumeration} the sets are consecutive. Otherwise, the time between two outputs could be too long. This theorem only gives us an output-sensitive results. Now, we prove that if we can bound $C_{\mathcal{P},G}$ polynomially, we can\longversion{ also} bound the input-sensitive running time.

\begin{lemma}\label{lem:nrdp_inherit_rdf_bounds}
    Let $\mathcal{P}$ be a \nrdp and $\mathcal{G}$ a class of graphs such that there is an algorithm enumerating all minimal \rdfs on $G=(V,E)\in \mathcal{G}$ in $\mathcal{O}^*(r^{\vert V\vert })$ for $r \in \{x\in \mathbb{R}\mid x\geq 1\}$.
    If there is an algorithm $\mathcal{A}$ that enumerates all elements in $C_{\mathcal{P},G}[A]$ with polynomial delay and if there is polynomial $p$ such that $\vert C_{\mathcal{P},G}[A]\vert \leq p(\vert V \vert)$ for each $G=(V,E) \in \mathcal{G}$ and $A\subseteq V$, then there is an algorithm to enumerate all minimal $g\in \{0,1,2\}^V$ with property~$\mathcal{P}$ on $G=(V,E)\in \mathcal{G}$ in $\mathcal{O}^*( r^{\vert V\vert })$.
\end{lemma}

\begin{pf}
    Let $G=(V,E)\in \mathcal{G}$ be a graph. 
    By \autoref{thm:nrdp_polyenumeration} there exists a polynomial-delay \longversion{enumeration }algorithm to enumerate all minimal $g\in \{0,1,2\}^V$ with property $\mathcal{P}$ on $G=(V,E)\in \mathcal{G}$. By \cite[Proposition 8]{AbuFerMan2024},\longversion{ it is known that} a minimal \rdf is uniquely defined by the vertices of value 2. Together with \autoref{lem:bij_min_rdf_set} there is a bijection between all minimal \rdfs on $G$ and subsets $A\subseteq V$ such that for all $v\in A$, $P_{G,A}(v) \nsubseteq \{v\}$. \shortversion{Hence}\longversion{Therefore}, the number of elements with property~$\mathcal{P}$ on~$G$ is bounded by $r^{\vert V\vert}\cdot p(\vert V\vert)$ for the given polynomial~$p$. As the delay is bounded by a polynomial~$q$, the running time of the algorithm is $q(\vert V \vert)p(\vert V \vert)r^{\vert V \vert}$ or $\mathcal{O}^*(r^n)$. 
\end{pf}

\section{Maximal Roman Dominating Functions}\label{sec:enum_mrdf}

In this section, we want to make use of the results in \autoref{sec:nrdp} by looking at maximal Roman dominating functions. For a graph $G=(V,E)$, a function $f \in \{0,1,2\}^V$ is a \emph{maximal Roman dominating function} (or mRdf for short) on $G$ \iffl $ N(v) \cap V_2(f) \neq \emptyset$ for all $v\in V_0(f)$ and $N[V_0(f)]\neq V$; see~\cite{ShaSLDJ2024}. 
First, we will show a characterization result for minimal \mrdfs.

\begin{theorem}\label{thm:property_minmaxrdf}\shortversion{$(\star)$}
    Let $G=(V,E)$ be a graph and $f\in \{0,1,2\}^V$. $f$ is a minimal \mrdf \iffl the following constraints hold.
    \begin{enumerate}
        \item $f$ is a\shortversion{n} \mrdf.\label{con:maxrdf_rdf}
        \item For all $v \in V_1(f)$, either $N(v) \cap V_2(f) = \emptyset$ or $V\setminus N[V_0(f)]\subseteq N[v]$.\label{con:maxrdf_V1}
        \item For all $v\in V_2(f)$, $P_{G[V_0(f) \cup V_2(f)],V_2(f)}(v)\nsubseteq \{v\}$.\label{con:maxrdf_private_neigh}
    \end{enumerate}
\end{theorem}
\begin{toappendix}
\begin{pf}\shortversion{(of \autoref{thm:property_minmaxrdf})}
    First we assume $f$ is a minimal \mrdf. Constraint~\ref{con:maxrdf_rdf} holds trivially. 
    Assume there exists a $v \in V_2(f)$ such that $v$ has no private neighbor but itself in $V_0(f) \cup V_2(f)$. Define $g \coloneqq f-\chi_{\{ v \}}$. Since $V_0(f) = V_0(g)$, $V_0(g)$ is no dominating set. So let $u \in V_0(f)$. If $u \notin N(v)$ then $u$ is dominated by $g$ as by $f$. For $u\in N(v)$, there exists a $w\in N(u) \cap (V_2(f) \setminus \{v\}) = N(u) \cap V_2(g)$. Hence, $g$ is a\shortversion{n} \mrdf. Thus, $f$ would not be minimal. This proves Constraint~\ref{con:maxrdf_private_neigh}.

    Let $v\in V_1(f)$. Assume Constraint~\ref{con:maxrdf_V1} does not hold for $v$. Define $g \coloneqq f-\chi_{\{ v \}}$. By assumption, $N(v)\cap V_2(g) = N(v) \cap V_2(f) \neq \emptyset$. As $N(u) \cap V_2(g) = N(u) \cap V_2(f) \neq \emptyset $ for $u\in V_0(f) = V_0(g) \setminus \{v\}$, $g$ is a \rdf.
    Furthermore, $V \setminus N [V_0(f)] \nsubseteq  N [v]$ implies $V\setminus N [V_0(g)] =  (V\setminus N[V_0(f)])\setminus N[v]\neq \emptyset$. Hence, $g$ is an \mrdf and $f$ was not minimal in the first place. 

    Conversely, assume $f$ fulfills all conditions. By Constraint~\ref{con:maxrdf_rdf}, $f$ is a\shortversion{n} \mrdf.  So assume there is a\shortversion{n} \mrdf $g\in \{0,1,2\}^V$ on $G$ with $g\leq f$ and $g\neq f$. Thus, $V_0(f)\subseteq V_0(g)$, $V_2(g) \subseteq V_2(f)$ and $V_1(g) \cup V_2(g) \subseteq V_1(f) \cup V_1(g)$. Let $v\in V$ be a vertex with $g(v)<f(v)$. First we assume $f(v)=2$. By Constraint~\ref{con:maxrdf_private_neigh}, there is a $u\in N(v) \cap V_0(f) \subseteq N(v) \cap V_0(g)$ such that $N(u) \cap V_2(g) \subseteq  N(u) \cap (V_2(f) \setminus \{v\}) = \emptyset$. This would imply that $g$ is no (maximal) \rdf.
    Hence, $f(v)=1$. If $\emptyset = N(v) \cap V_2(f) (\subseteq N(v) \cap V_2(g))$, $g$ is no \rdf as $g(v)=0$. Therefore, $V\setminus N[V_0(f)] \subseteq N[v]$. Then, $V\setminus N[V_0(g)] = \emptyset$  and $V_0(f)$ is a \ds on $G$. In this case $g$ would not be a\shortversion{n} \mrdf.   
\end{pf}
\end{toappendix}

\longversion{\begin{corollary}\label{cor:max_rdf_V1_not_dom}\shortversion{$(\star)$}
    For a graph $G=(V,E)$ and a minimal \mrdf $f\in \{0,1,2\}^V$ on~$G$, $\emptyset \neq V \setminus N[V_0(f)] \subseteq V_1(f)$.
\end{corollary}

\begin{toappendix}
\begin{pf}\shortversion{(of \autoref{cor:max_rdf_V1_not_dom})}
    As $f$ is a\shortversion{n} \mrdf, there exists a $v\in V \setminus N[V_0(f)]$. If $f(v) = 2$, then $v$ would not have a private neighbor but itself,  contradicting Constraint~\ref{con:maxrdf_private_neigh}. 
\end{pf}
\end{toappendix}}

To show that our approach is really helpful, we prove hardness for the extension problem. First, let us consider a non-trivial no-instance.  Let $G=(\{1,2,3,4\},\{\{1,2\},\{2,3\},\{3,4\}\})$  be a path of~4 vertices with the function $f\coloneqq 2\cdot\chi_{\{1,4\}}$. This instance is visualized by \autoref{fig:no-instance_emrd}. $f$ is a \rdf, but $V_0(f)$ is a \ds. Furthermore, each vertex in $V_0(f)$ is a unique private neighbor of a vertex in $V_2(f)$. \autoref{thm:property_minmaxrdf} implies, that for any $g\in \{0,1,2\}^{\{1,2,3,4\}}$ with $f\leq g$, $g$ is not a minimal \mrdf.  
\begin{figure}[bt]
    \centering    	
	\begin{tikzpicture}[transform shape]
			\tikzset{every node/.style={ fill = white,circle,minimum size=0.3cm}}

			\node[draw] (a1) at (-2,0) {2};
			\node[draw] (a2) at (0,0) {0};
			\node[draw] (a3) at (2,0) {0};
			\node[draw] (a4) at (4,0) {2};
   
            \path (a1) edge[-] (a2);
            \path (a2) edge[-] (a3);
            \path (a3) edge[-] (a4);
        \end{tikzpicture}

    \caption{Non-trivial no-instance of \EMRD.}
    \label{fig:no-instance_emrd}
\end{figure}
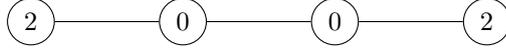
\begin{theorem}\label{thm:Ext_maxRdf_NP}\shortversion{$(\star)$}
    \EMRD is \NP-complete even when $\vert V_1(f)\vert =2$\longversion{ ($\,\vert V_1(f)\vert$-\EMRD is \paraNP-complete)}. \longversion{$\vert V_2(f)\vert$-\EMRD is \W{3}-hard (even on bipartite graphs).}
\end{theorem}
\begin{toappendix}
    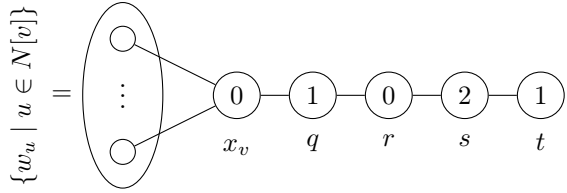
\begin{figure}[bt]
    \centering
    	    	
	\begin{tikzpicture}[transform shape]
			\tikzset{every node/.style={ fill = white,circle,minimum size=0.3cm}}
            \node  at (-2.8,0)[rectangle,label={[rotate=90]center:$\{w_u\mid u\in N[v]\}$}]{};
            \node [] at (-2.3,0){$=$};
            \node at (-1.5,0.1) {\vdots};	
            \draw (-1.5,0) ellipse (15pt and 35pt);	

			\node[draw,label={below:$x_v$}] (v2) at (0,0) {0};
			\node[draw,label={below:$q$}] (a) at (1,0) {1};
			\node[draw,label={below:$r$}] (c) at (2,0) {0};
			\node[draw,label={below:$s$}] (d) at (3,0) {2};
			\node[draw,label={below:$t$}] (e) at (4,0) {1};
			\node[draw] (vk11) at (-1.5,0.75) {};
			\node[draw] (vk12) at (-1.5,-0.75) {};
   
            \path (v2) edge[-] (vk11);	
            \path (v2) edge[-] (vk12);	
            \path (v2) edge[-] (a);				            
            \path (a) edge[-] (c);            
            \path (c) edge[-] (d);            
            \path (d) edge[-] (e);     
        \end{tikzpicture}

    \label{fig:paraNP_ERD_V1_neighbor_V2}
    \caption{Construction for \autoref{thm:Ext_maxRdf_NP} with $v\in V$.}
    \end{figure}
\begin{pf}\shortversion{(of \autoref{thm:Ext_maxRdf_NP})}
    \EMRD is in \NP, as the Turing machine can guess the values of~$g$ and then use \autoref{thm:property_minmaxrdf} for polynomial-time verification.
    We will prove the hardness results by a reduction from \EDS. Let $G=(V,E)$ be a graph and $U\subseteq V$ a vertex set. Define $\widetilde{G}=(\widetilde{V},\widetilde{E})$ with 
    
    \begin{equation*}
        \begin{split}
            \widetilde{V} \coloneqq{}& \{q,r,s,t\} \cup \{w_v, x_v, \mid v\in V\} \\
            \widetilde{E} \coloneqq{}& \{\{w_v, x_u\}\mid v,u\in V, u\in N[v]\} \cup{}\\ & \{\{x_v,q\} \mid v\in V\} \cup \{\{q, r\}, \{r,s\}, \{s,t\}\}.
        \end{split}
    \end{equation*}   
    $\widetilde{G}$ is bipartite with the partition $ \{q,s\} \cup \{w_v, \mid v\in V\}, \{r,t\} \cup \{ x_v \mid v\in V\}$. Furthermore, define $f\in \{0,1,2\}^{\widetilde{V}}$ with $$f=2\chi_{\{s\}\cup \{w_v \mid v\in U\}} + \chi_{\{q,t\}}.$$
    \begin{claim}
        There exists a minimal \ds $D$ on $G$ with $U \subseteq D$ \iffl there is minimal \mrdf $g$ on $\widetilde{G}$ with $f\leq g$.  
    \end{claim}
    \begin{pfclaim} 
        Let $D$ be a minimal dominating set (\ds) on $G=(V,E)$ with $U \subseteq D$. Define $g_D=f+ \chi_{\{w_v\mid v\in D \setminus U\}}$. Clearly, $f \leq g_D$. Vertex~$s$ is a neighbor of $r$. As $D$ is a \ds, for each $v\in V$, $x_v$ has a neighbor $w_v\in \{ w_u \mid u \in V\} \cap V_2(f)$. Thus, $g_D$ is a \rdf. Since $N[t] = \{s,t\} \subseteq V_1(g_D) \cup V_2(g_D)$, $g_D$ is a\shortversion{n} \mrdf. As each $v\in D$ has a private neighbor, for every $w_v\in V_2(f) \cap \{w_x\mid x\in V\}=\{w_v\mid v\in D\}$, there exists a private neighbor $x_u\in \{x_p \mid p \in V \}$. Furthermore, $r$ is a private neighbor of $s$. Since $V_1(g_D) \cap N(V_2(g_D)) =\{t\}$ and $N(t)=\{s\}$, \autoref{thm:property_minmaxrdf} implies that $g_D$ is a minimal \mrdf. 

        Now let $g$ be a minimal \mrdf with $f\leq g$. Since $r$ is the only neighbor of~$s$ in $V_0(f)$, $g(r)=0$ and $V_2(g)\cap  N(r) = \{s\}$. Thus, $g(q)=1$.
        If there is a $v\in V$ with $g(x_v) = 2$, $q$ would contradict Constraint~\ref{con:maxrdf_V1} of \autoref{thm:property_minmaxrdf}. Hence, for each $v\in V$, $N(w_v)\cap V_2(g) = \emptyset$ and $g(w_v)\neq 0$. For all $v\in V$, $N(x_v)\subseteq V_1(g)\cap V_2(g)$ which implies $\{x_v\mid v\in V\} \subseteq V_0(g)$. Otherwise, $t$ would contradict Constraint~\ref{con:maxrdf_V1} of \autoref{thm:property_minmaxrdf}.
         Therefore, for each $v\in V$ there exists a $u\in V$ such that $w_u\in V_2(g)\cap N(x_v)$. Thus, $D_g \coloneqq \{v\in V\mid w_v\in V_2(g)\}$ is a dominating set. Let $v\in D_g$. Since $g$ is minimal, $w_v$ has a private neighbor~$x_u$. Hence, $v$ has a private neighbor and $D_g$ is minimal. 
    \end{pfclaim}

    The graph $\widetilde{G}$ has $2 \vert V\vert +4$ vertices. Therefore, the reduction can be implemented to run in polynomial time. \longversion{Since $\vert V_2(f)\vert = \vert U\vert + 1$, this is also a \FPT-reduction.}
\end{pf}
\end{toappendix}

Let $\mathcal{MRDF}$ denote the property of \mrdfs. To apply earlier results, we now show that  $\mathcal{MRDF}$ is nice.

\begin{lemma}\label{lem:MRDF_nrdp}
    $\mathcal{MRDF}$ is a \nrdp.
\end{lemma}

\begin{pf}
    Clearly, every \mrdf is also a \rdf. Let $G=(V,E)$ be a graph and let $f$ be a\shortversion{n} \mrdf on $G$. Since domination and Roman domination are monotonically increasing (with respect to $\subseteq$ and $\leq$), $f+\chi_{\{v\}}$ is still a\shortversion{n} \mrdf for all $v \in V_0(f)$. Therefore, we only need to consider the Constraint \ref{con:nrdp_V_2}.

    Let $v\in V_2(f)$ and $f_v \coloneqq f-\chi_{\{v\}}$. Clearly, $V_0(f_v)=V_0(f)$. Therefore, if $f_v$ is no \mrdf, there is a $u\in V_0(f_v)$ which is not dominated by $V_2(f_v)=V_2(f)\setminus \{v\}$. Hence, $u\in N(v)$ is a private neighbor of~$v$. 
    If $P_{G[V\setminus V_1(f)],V_2\left(f\right)}\left( v \right) \subseteq \lbrace v\rbrace$, then for each $w \in V_0(f)=V_0(f_v)$, there exists a $u\in (V_2(f)\setminus \{v\}) \cap N(w)= V_2(f_v) \cap N(w)$. Therefore, $V_0(f)=V_0(f_v)$ is dominated by $V_2(f)\setminus \{ v\} = V_2(f_v)$.
\end{pf}

\begin{lemma}\label{lem:bounded_CV_2}
    Let $G=(V,E)$ be a graph and let $A \subseteq V$. Then, $\vert C_{\mathcal{MRDF},G}[A]\vert \leq n$. Furthermore, $C_{\mathcal{MRDF},G}[A]$ can be computed in polynomial time.
\end{lemma}

\begin{pf}
We can assume that each $v\in A$ has a private neighbor but itself. Otherwise, $C_{\mathcal{MRDF},G}[A]=\emptyset$ by Constraint \ref{con:maxrdf_private_neigh} of \autoref{thm:property_minmaxrdf}. This property can be checked in polynomial time.

Define $f\in \{0, 1, 2\}^V$ with $f \coloneqq 2 \cdot \chi_{A} + \chi_{V\setminus N[A]}$. As $V_0(f) \subseteq N(A) = N(V_2(f))$, $f$ is a (minimal) \rdf. Since each $v \in V\setminus N[A]$ has no neighbor in $V_2$, $f \leq g$ for all $g\in C_{\mathcal{MRDF},G}[A]$. Therefore, if $f$ is a\shortversion{n} \mrdf, then $C_{\mathcal{MRDF},G}[A]=\{f\}$. So, assume there is no $v\in V_1(f)$ such that \shortversion{$N(v) \cap V_0(f)\neq \emptyset$}\longversion{$N(v) \cap V_0(f)$ is not empty}. Then define for all $v\in V\setminus A$, $g_v\in \{0,1,2\}^V$ with $g_v=f+ \chi_{V_0(f) \cap N[v]}$. Clearly, $f\leq g_v$ and $g_v$ is a \rdf. Since $N[v] \cap V_0(g_v) = N[v] \cap (V_0(f)\setminus N[v])$, $g_v$ is a\shortversion{n} \mrdf.

Assume there is a minimal \mrdf $h$ on $G$ such that $V_2(h)=A$, but there is no $v\in V\setminus A$ with $h=g_v$. As $h$ is a \rdf, $f\leq h$. By \shortversion{\autoref{thm:property_minmaxrdf}}\longversion{\autoref{cor:max_rdf_V1_not_dom}}, there is a $v \in V_1(h)\subseteq V\setminus A$, such that $v\notin N[V_0(h)]$. Hence, for each $v\in V\setminus A$, $g_v= f + \chi_{N[v]\cap V_0(f)} \leq h$. Thus, $h$ is not minimal in the first place and $C_{\mathcal{MRDF},G}[A] \subseteq \{g_v\mid v\in V\setminus A\}$. Therefore, $\vert C_{\mathcal{MRDF},G}[A]\vert \leq \vert V\setminus A\vert\leq n$.

It can be checked in polynomial time if $f$ and $g_v$ (for all $v\in V \setminus V_2$) are minimal \mrdfs. \shortversion{Thus}\longversion{This implies that} $C_{\mathcal{MRDF},G}[A]$ can be constructed in polynomial time. 
\end{pf}

\noindent
\autoref{thm:nrdp_polyenumeration} and \autoref{lem:nrdp_inherit_rdf_bounds} imply with Theorem 11 of \cite{AbuFerMan2024} the following:

\begin{corollary}There is a polynomial-space algorithm that enumerates all minimal \mrdfs of a given graph of order $n$ with polynomial delay, in time $\mathcal{O}(1.9332^n)$.
\end{corollary}
\noindent
The results in \cite{AbuFerMan2023} imply the following:
\begin{corollary}
    There is a polynomial-space algorithm that enumerates all minimal \mrdfs of a given graph of order $n$ with polynomial delay and in time
    \begin{itemize}
        \item $\mathcal{O}(1.8940^n)$ on chordal graphs;
        \item $\mathcal{O}(\sqrt{3}^n)$ on interval graphs or on trees;
        \item $\mathcal{O}(\sqrt[3]{3}^n)$ on split or on cobipartite graphs.
    \end{itemize}
\end{corollary}

The next corollary follows by \autoref{cor:Ext_nrdp_1_2_notneighbor} and \autoref{lem:bounded_CV_2}.

\begin{corollary}\label{cor:Ext_mrdf_1_2_notneighbor}
    \EMRD is polynomial time solvable for the instance $(G,f)$ if $N(V_1(f))\cap V_2(f)=\emptyset$.
\end{corollary}

\begin{remark}\label{rem:strength_nrdp}
    In the proof of \autoref{thm:Ext_maxRdf_NP} we construct a graph $\widetilde{G}=(\widetilde{V},\widetilde{E})$.
    For the hardness we makes use of a $t\in \widetilde{V}$ of value~1 for which a $s\in \widetilde{V}$ of value~2 is the only neighbor. Because of Constraint~\ref{con:maxrdf_V1} of \autoref{thm:property_minmaxrdf}, there is no $w\in \widetilde{V} \setminus \{t\}$ of value~1 without a neighbor of value~0. 
    
    This is interesting since the instance  would be polynomial time solvable, if the value of $t$ would be~0 (see \autoref{cor:Ext_mrdf_1_2_notneighbor}). Therefore, just one wrongly placed~1 is enough to make the extension problem hard. This is the motivation for considering the vertices of value~1 after choosing the vertices with the value~2.  
\end{remark}

\section{Extension Total/Connected Roman Domination}\label{sec:enum_tcrdf}

For this chapter we want to consider total and connected Roman Domination. The so-called total Roman Domination parameter was introduced in \cite{LiuCha2013}. A \rdf $f\in\{0,1,2\}^V$ is a \emph{total Roman dominating function} (or \emph{tRdf} for short) with respect to $G$ if, for each $v\in V_1(f)\cup V_2(f)$, $N(v)\cap\left( V_1(f)\cup V_2(f)\right)$ is not empty.  A \emph{connected Roman dominating function} (or \emph{cRdf} for short) with respect to $G$ is a \rdf such that $ V_1(f)\cup V_2(f)$ is connected in $G$.\footnote{There is a different definition in \cite{MudSum2013}. In our opinion, the given definition fits better than the original, military motivation of distributing legions.}
References~\cite{LiRZD2022,LiuCha2013} provide proofs for the \NP-completeness for the natural decision problem of \textsc{Total} or \textsc{Connected Roman Domination} (given a graph $G=(V,E)$ and $k\in \mathbb{N}$, is there a \trdf or \crdf $f$ with $\sum_{v\in V}f(v)\leq k$).
We start by proving a characterization result of a minimal \trdf/\crdf:
\shortversion{
\begin{theorem}\label{thm:minimal_tcrdf}\shortversion{$(\star)$}
      Let $G=(V,E)$ be a graph and $f\in \{0,1,2\}^V$. $f$ is a minimal \trdf (or \crdf, resp.) \iffl the following conditions hold:
\begin{enumerate}
    \item $f$ is a \trdf  (or \crdf, resp.),\label{con_tcrdf_ground}
    \item for all $v\in V_1(f)$, (a) $N(v)\cap V_2(f)=\emptyset$, or (b) $G[(V_1(f) \cup V_2(f))\setminus \{v\}]$ has an isolated vertex (or (b)  $G[(V_1(f) \cup V_2(f))\setminus \{v\}]$ is not connected, resp.) and \label{con_tcrdf_V1}
    \item for all $ v\in V_2\left(f\right)$, $ P_{G[V_0(f)\cup V_2(f)],V_2\left(f\right)}\left( v \right) \nsubseteq \lbrace v\rbrace$. \label{con_tcrdf_private}
\end{enumerate}
\end{theorem}}

\begin{toappendix}
\shortversion{In order to prove \autoref{thm:minimal_tcrdf}, we formulate and show two lemmata.}

\begin{lemma}\label{lem:minimal_trdf}
    Let $G=(V,E)$ be a graph and $f\in \{0,1,2\}^V$. $f$ is a minimal  \trdf \iffl the following conditions hold:
\begin{enumerate}
    \item $f$ is a \trdf,\label{con_trdf_ground}
    \item for all $v\in V_1(f)$, $N(v)\cap V_2(f)=\emptyset$ or $G[(V_1(f) \cup V_2(f))\setminus \{v\}]$ has an isolated vertex and \label{con_trdf_V1}
    \item for all $ v\in V_2\left(f\right)$, $ P_{G',V_2\left(f\right)}\left( v \right) \nsubseteq \lbrace v\rbrace$. \label{con_trdf_private}
\end{enumerate}
\end{lemma}
\begin{pf}
    First let $f$ be a minimal \trdf. Hence, $f$ is a \trdf. Assume there is a $v \in V_1(f)$ where $N(v)\cap V_2(f) \neq \emptyset$ and $G[(V_1(f) \cup V_2(f))\setminus \{v\}]$ has no isolate. Define $g\coloneqq f- \chi_{\{v\}}$. Clearly, each $x \in V_0(f)=V_0(g)\setminus \{v\}$ is dominated by $g$ as it is by $f$. Since $N(v)\cap V_2(f) \neq \emptyset$, $v$ is also dominated and $g$ a \rdf. As $G[V_1(g)\cup V_2(g)] = G[(V_1(f) \cup V_2(f))\setminus \{v\}] $ has no isolates, $g$ is a \trdf. 

    Now assume there is a $v\in V_2(f)$ such that $P_{G',V_2\left(f\right)}\left( v \right) \subseteq \lbrace v\rbrace$. Define $g\coloneqq  f - \chi_{\{v\}}$. If $u$ is a neighbor of~$v$, there exists a vertex $w\in N_{G}\left[u\right] \cap\left( V_2\left(f\right) \setminus \lbrace v\rbrace\right) = N_{G}\left[u\right]\cap V_2\left(g\right)$. For $u\in V_0\left(g\right)$ with $u\notin N(v)$, there is a vertex $w\in V_2\left(f\right)\setminus\{v\}$ that dominates~$u$. Hence, the three properties hold. 

    For the other implication, let the three properties be valid for $f$. Let $h$ be a minimal \trdf with $h \leq f$. This implies $V_2(h)\subseteq V_2(f)$.
    In the following we assume there is a $v\in V$ such that $h(v) < f(v)$  and go through all the cases.
    
    \textbf{Case 1: $f(v)=2$.} By Property \ref{con_trdf_private}, there is a $u\in (N(v) \cap V_0(f))\setminus N(V_2(f)\setminus \{v\})\subseteq (N(v) \cap V_0(h))\setminus N(V_2(h))$. Since $v\notin V_2(h)$, $u$ is not dominated. This would contradict that $g$ is a \rdf. Hence, we can assume $V_2(f)=V_2(h)$. This implies $V_1(h)\subseteq V_1(f)$.

    \textbf{Case 2: $h(v)=0<1=f(v)$.} Property \ref{con_trdf_V1} implies that $N(v) \cap V_2(f)= N(v) \cap V_2(h) =\emptyset$ or $G[(V_1(f) \cup V_2(f))\setminus \{v\}]$ has an isolate. If the first part would hold, $v$ would not be dominated by $h$ which is a contradiction. Therefore, we can assume there is a $u\in N(v)\cap V_2(f)$ and $G[(V_1(f) \cup V_2(f))\setminus \{v\}]$ has isolates. As $f$ is a \trdf, the isolated vertex $x$ is in $N(v)$. If $x\in V_2(f)=V_2(h)$, $x$ would be isolated in $G[V_1(h)\cup V_2(h)]$, since $V_1(h)\subseteq V_1(f)$ and $V_2(f)=V_2(h)$. This would contradict the totality of $h$. Hence, $x\in V_1(f)$. Since $x$ is an isolate in $G[(V_1(f)\cup V_2(f))\setminus \{v\}]$, $x$ has no neighbor in $V_2(f)$. Thus, under condition $h(x)=0$, $h$ would not be a \rdf. So, $h(x)=1$ which implies that $x$ is an isolate on the subgraph $G[V_1(h) \cup V_2(h)]$ of $G[(V_1(f)\cup V_2(f))\setminus \{v\}]$. This would contradict the totality of $h$. 
    Hence the equivalence holds.
\end{pf}

\begin{lemma}
\label{lem:minimal_crdf}
    Let $G=(V,E)$ be a graph and $f\in \{0,1,2\}^V$. $f$ is a minimal  \crdf \iffl the following conditions hold:
\begin{enumerate}
    \item $f$ is a \crdf,\label{con_crdf_ground}
    \item for all $v\in V_1(f)$, $N(v)\cap V_2(f)=\emptyset$ or $G[(V_1(f) \cup V_2(f))\setminus \{v\}]$ is not connected and \label{con_crdf_V1}
    \item for all $v\in V_2\left(f\right)$, $ P_{G',V_2\left(f\right)}\left( v \right) \nsubseteq \lbrace v\rbrace$. \label{con_crdf_private}
\end{enumerate}
\end{lemma}
    \begin{pf}
    First let $f$ is a minimal \crdf. Hence, $f$ is a \crdf. Assume there is a $v \in V_1(f)$ where $N(v)\cap V_2(f) \neq \emptyset$ and $G[(V_1(f) \cup V_2(f))\setminus \{v\}]$ is connected. Define $g\coloneqq f- \chi_{\{v\}}$. Clearly, each $x \in V_0(f)=V_0(g)\setminus \{v\}$ is dominated by $g$ as it is by $f$. Since $N(v)\cap V_2(f) \neq \emptyset$, $v$ is also dominated and $g$ a \rdf. As $G[V_1(g)\cup V_2(g)] = G[(V_1(f) \cup V_2(f))\setminus \{v\}] $ is connected, $g$ is a \crdf. 

    Now, assume there is a $v\in V_2(f)$ such that $P_{G',V_2\left(f\right)}\left( v \right) \subseteq \lbrace v\rbrace$. Define $g\coloneqq  f - \chi_{\{v\}}$. If $u$ is a neighbor of~$v$, there exists a vertex $w\in N_{G}\left[u\right] \cap\left( V_2\left(f\right) \setminus \lbrace v\rbrace\right) = N_{G}\left[u\right]\cap V_2\left(g\right)$. For $u\in V_0\left(g\right)$ with $u\notin N(v)$, there is a vertex $w\in V_2\left(f\right)\setminus\{v\}$ that dominates~$u$. Furthermore, $V_1(f)\cup V_2(f)=V_1(g)\cup V_2(g)$ is connected, which implies that $g$ is a \crdf. Hence, the three properties hold. 

    For the other implication, let the three properties be valid for $f$. Let $h$ be a minimal \crdf with $h \leq f$. This implies $V_2(h)\subseteq V_2(f)$.
    In the following we assume there is a $v\in V$ such that $h(v) < f(v)$  and go through all the cases.
    
    \textbf{Case 1: $f(v)=2$.} By Property \ref{con_crdf_private}, there is a $u\in (N(v) \cap V_0(f))\setminus N(V_2(f)\setminus \{v\})\subseteq (N(v) \cap V_0(h))\setminus N(V_2(h))$. Since $v\notin V_2(h)$, $u$ is not dominated. This would contradict that $g$ is a \rdf. Hence, we can assume $V_2(f)=V_2(h)$. This implies $V_1(h)\subseteq V_1(f)$.

    \textbf{Case 2: $h(v)=0<1=f(v)$.} Property \ref{con_crdf_V1} implies that $N(v) \cap V_2(f)= N(v) \cap V_2(h) =\emptyset$ or $G[(V_1(f) \cup V_2(f))\setminus \{v\}]$ is not connected. If the first part would hold, $v$ would not be dominated by $h$ which is a contradiction. Therefore, we can assume there is a $u\in N(v)\cap V_2(f)$ and $G[(V_1(f) \cup V_2(f))\setminus \{v\}]$ is not connected. $h \leq f$ implies $\overline{(V_1(f)\cup V_2(f))} = V_0(f) \subseteq  V_0(h)=\overline{(V_1(h)\cup V_2(h))}$. Thus, $V_1(h)\cup V_2(h) \subseteq V_1(f)\cup V_2(f)$. So, if $V_1(f)\cup V_2(f)$ is not connected then $V_1(h)\cup V_2(h)$ is neither.
    Hence the equivalence holds.
\end{pf}
\end{toappendix}

\noindent
This result is important for the remaining section, especially for the next result:

\begin{theorem}\shortversion{$(\star)$}\label{thm:CTRDF_nrdp}
    $\mathcal{CRDF}$ and $\mathcal{TRDF}$ are \nrdps.
\end{theorem}
\begin{toappendix}
  
\begin{pf}\shortversion{(of \autoref{thm:CTRDF_nrdp})}
    Clearly, every \crdf (resp. \trdf) is also a \rdf. Let $G=(V,E)$ be a graph and $f$ a \crdf (resp. \trdf) on $G$. Since $f$ is \rdf, $V_0(f)\subseteq N(V_1(f)\cup V_2(f))$. Hence, for all $v\in V_0(f)$, $\{v\}\cup V_1(f)\cup V_2(f)$ is connected (resp. has no isolated vertex). This implies, $f+\chi_{\{v\}}$ is still a \crdf (resp. \trdf) for all $v \in V_0(f)$. Therefore, we only need to consider the Constraint \ref{con:nrdp_V_2}.

    Let $v\in V_2(f)$ and $f_v \coloneqq f-\chi_{\{v\}}$. Clearly, $V_1(f_v)\cup V_2(f_v)=V_1(f) \cup V_2(f)$. Therefore, if $f_v$ is no \crdf (resp \trdf), there is a $u\in V_0(f_v)$ which is not dominated by $V_2(f_V)=V_2(f)\setminus \{v\}$. Hence, $u\in N(v)$ is a private neighbor of $v$. 
    If $P_{G[V\setminus V_1(f)],V_2\left(f\right)}\left( v \right) \subseteq \lbrace v\rbrace$, then for each $w \in V_0(f)=V_0(f_v)$ there exists a $u\in (V_2(f)\setminus \{v\}) \cap N(w)= V_2(f_v) \cap N(w)$. Therefore, $V_0(f)=V_0(f_v)$ is dominated by $V_2(f)\setminus \{ v\} = V_2(f_v)$.
\end{pf}
\end{toappendix}

\shortversion{\subsection{Minimal \crdf/\trdf on Cobipartite Graphs}}
\longversion{\subsection{Minimal connected/\trdf on Cobipartite Graphs}}

A graph $G=(V,E)$ is called \emph{cobipartite} if there is a partition of $V$ into two cliques $C_1,C_2$.  It is known by \cite{KobKMMO2025} that enumerating minimal connected dominating sets is as hard as enumerating minimal hitting sets. The goal of this subsection is the following theorem:  

\begin{theorem}\label{thm:enum_tcrdf_on_cobi}\shortversion{$(\star)$}
For all cobipartite graphs $G=(V,E)$  and $A\subseteq V$, $C_{\mathcal{CRDF},G}[A]$ and $C_{\mathcal{TRDF},G}[A]$ can be enumerated with polynomial delay and $\vert C_{\mathcal{CRDF},G}[A]\vert\in \mathcal{O}(\vert V \vert^2)$ and $ \vert C_{\mathcal{TRDF},G}[A]\vert \in \mathcal{O}(\vert V \vert^2)$. 
Furthermore, there is a polynomial-delay algorithm enumerating all minimal \crdfs (or \trdfs) on cobipartite  graphs of order~$n$ within time $\Oh^*(\sqrt[3]3^n)$.    
\end{theorem}

We will shortly describe the idea of the proof. \shortversion{Details can be found in \todo{arxiv}.} Let $G=(V,E)$ be a cobipartite graph with the two cliques $C_1,C_2$, $A\subseteq V$, $f_A \coloneqq 2\cdot \chi_A + \chi_{V\setminus N[A]}$ and $g\in C_{\mathcal{XRDF},G}[A]$, where $\mathcal{X}\in \{\mathcal{T,C}\}$. We can conclude $\vert (V_1(g) \cap C_i) \setminus V_1(f_A)\vert\leq 1$ for $i\in \{1,2\}$. Therefore, $\vert C_{\mathcal{XRDF},G}[A] \vert \leq \mathcal{O}(\vert V\vert^2 )$ and we can enumerate all elements in $C_{\mathcal{XRDF},G}[A]$ with polynomial delay. 

\begin{toappendix}

\shortversion{\subsection{Proof of \autoref{thm:enum_tcrdf_on_cobi}}}
Now we want to prove \autoref{thm:enum_tcrdf_on_cobi}. For this we will first consider some lemmas.

\begin{lemma}\label{lem:crdf_V_1_co-bi}
    Let $G=(V,E)$ be cobipartite graph where $C_1,C_2$ are two disjoint cliques with $V=C_1 \cup C_2$. For all minimal \crdfs $f\in \{0,1,2\}^V$ on $G$, $\vert V_1(f) \cap N(V_2(f)\cap C_1)\vert \leq 1$ and $\vert V_1(f) \cap N(V_2(f)\cap C_2)\vert \leq 1$.
\end{lemma}
    \begin{pf}
    Let $T \coloneqq V_2(f)\cap C_1$ and $v,u\in N(T)\cap V_1(f)$. Denote some neighbor of $v$ and $u$ in $T$ by $x_v$ and $x_u$. Since $v,u$ are dominated by $T$, $(V_1(f) \cup V_2(f))\setminus \{v\}$ and $(V_1(f) \cup V_2(f))\setminus \{u\}$ are not connected. As $C_1,C_2$ are cliques, $C_1\cap (V_1(f) \cup V_2(f))$ is connected. Hence, there is a $y\in (V_1(f)\cup V_2(f)) \cap C_2 \cap ( N(u)\setminus N( T \cup \{v\}))$. Thus, for all $z\in N(v)\setminus N(\{u,x_u,x_v\})\subseteq C_2$, $x_v,x_u,u,y,z$ is a path from $x_v$ to~$z$. Therefore, $(V_1(f) \cup V_2(f))\setminus \{v\}$ is connected and $f-\chi_{\{v\}}$ a \crdf.  
\end{pf}

\begin{lemma}\label{lem:crdf_cobi_C_poly_bound}
    Let $G=(V,E)$ be a cobipartite graph of order $n$ and $A\subseteq V$. Then $\vert C_{\mathcal{CRDF},G}[A]\vert \in \mathcal{O}(n^2)$ and $C_{\mathcal{CRDF},G}[A]$ can be enumerated in polynomial time.
\end{lemma}

\begin{pf}
    Let $C_1,C_2$ be disjoint cliques with $C_1\cup C_2=V$. By \autoref{lem:NRDP_G_eq_extension}, for each $g\in C_{\mathcal{CRDF},G}[A]$, $f_A \coloneqq 2 \cdot \chi_A + \chi_{V\setminus N[A]} \leq g$. If $A=\emptyset$, $f_A = \chi_{V}$ which is a minimal \crdf.
    
    Let $A\cap C_1 \neq \emptyset$ and $A\cap C_2=\emptyset$ (the other way around works analogously). If $V_1(f_A)= \emptyset$, then $f_A$ is a \crdf. Otherwise $f_A$ is no \crdf, as $V_0(f_A)$ separates $V_1(f_A)$ from $V_2(f_A)=A$. So we consider the case $V_1(f_A)\neq \emptyset$. By \autoref{lem:crdf_V_1_co-bi}, for all minimal \crdfs $g$ on $G$ with $V_2(g)=A$, $\vert V_1(g) \cap N(A\cap C_1)\vert \leq 1$. Thus, $C_{\mathcal{CRDF},G}[A] \subseteq \{f_A+\chi_{\{v\}} \mid v\in V_0(f)\}$. Since $\vert \{f_A+\chi_{\{v\}} \mid v\in V_0(f)\}\vert \leq n$ and checking if a function is a minimal \crdf can be done in polynomial time, we can enumerate all elements in polynomial time. 

    The only case we still need to consider is $A\cap C_1 \neq \emptyset$ and $A\cap C_2 \neq \emptyset$. In this case $A$ is a dominating set, which implies $V_1(f_A)= \emptyset$. Let $g$ be a minimal \crdf with $V_2(g)=A$. By \autoref{lem:crdf_V_1_co-bi}, \begin{equation*}
        \begin{split}
            \vert V_1(g)\vert ={}& \vert (V_1(g)\cap C_1)\cup (V_1(g)\cap C_2) \vert  
            \leq \vert V_1(g)\cap C_1\vert + \vert  V_1(g)\cap C_2 \vert\\
            \leq{}& \vert V_1(g) \cap N(A \cap C_2) \vert + \vert V_1(g) \cap N(A \cap C_2) \vert \leq 2
        \end{split}
    \end{equation*} 
    Thus, $C_{\mathcal{CRDF},G}[A] \subseteq  \{f_A \} \cup \{f_A+\chi_{\{v\}} \mid v\in V_0(f_A)\} \cup \{f_A+\chi_{\{v,u\}} \mid v,u\in V_0(f)\}$ which implies $\vert C_{\mathcal{CRDF},G}[A] \vert \leq n^2 + n + 1$. Since checking if a function is a minimal \crdf can be done in polynomial time, we can enumerate all elements in polynomial time. 
\end{pf}

\noindent
\autoref{lem:nrdp_inherit_rdf_bounds}, \autoref{thm:CTRDF_nrdp} and \autoref{lem:crdf_cobi_C_poly_bound} together with \cite{AbuFerMan2023} imply:

\begin{corollary}\label{cor:poly_delay_enum_crdf_cobi}
    There is a polynomial-delay algorithm enumerating all minimal \crdfs on cobipartite of order $n$ graphs within time $\Oh^*(\sqrt[3]3^n)$.
\end{corollary}

\noindent
With the same strategy, we can obtain similar results for \trdf\longversion{ on cobipartite graphs}: 

\begin{lemma}\label{lem:trdf_cobi_C_poly_bound}
    Let $G=(V,E)$ be a cobipartite graph of order $n$ and $A\subseteq V$. Then $\vert C_{\mathcal{TRDF},G}[A]\vert \in \mathcal{O}(n^2)$ and $C_{\mathcal{TRDF},G}[A]$ can be enumerated in polynomial time.
\end{lemma}
\begin{pf}
    Let $C_1,C_2$ be disjoint cliques with $C_1\cup C_2=V$. By \autoref{lem:NRDP_G_eq_extension}, for each $g\in C_{\mathcal{CRDF},G}[A]$, $f_A \coloneqq 2 \cdot \chi_A + \chi_{V\setminus N[A]} \leq g$. 
    \begin{claim}
        $\vert (V_1(g) \cap C_1) \setminus V_1(f_A)\vert \leq 1$
    \end{claim}

    \begin{pfclaim}
        Assume there are two $v,u \in (V_1(g) \cap C_1) \setminus V_1(f_A)$ with $v\neq u$. Since $g\geq f_A$ and $f_A$ is a \rdf, there are $x_v,x_u\in A$ such that $\{v,x_v\}, \{u,x_u\}\in E$. Thus, $(V_1(g)\cup V_2(g))\setminus \{u\}$ and $(V_1(g)\cup V_2(g))\setminus \{v\}$ have isolates. We will now consider $(V_1(g)\cup V_2(g))\setminus \{u\}$. Since $v\in C_1$ and $x_v\in N(v)$, there is no isolate in $C_1 \cap ((V_1(g)\cup V_2(g))\setminus \{u\})$. Hence, the isolate has to be in $C_2$. Thus, $\vert C_2 \cap ((V_1(g)\cup V_2(g))\setminus \{u\})\vert =1$. Let $y$ be this isolate. Therefore, $N(y)\cap (V_1(g) \cup V_2(g))=\{u\}$ and $N(v)\cap C_2=\emptyset$. This implies $(V_1(g)\cup V_2(g))\setminus \{v\}$ has no isolates which is a contradiction to the minimality of $g$.
    \end{pfclaim}
    
    Thus, $C_{\mathcal{TRDF},G}[A] \subseteq  \{f_A \} \cup \{f_A+\chi_{\{v\}} \mid v\in V_0(f_A)\} \cup \{f_A+\chi_{\{v,u\}} \mid v,u\in V_0(f)\}$ which implies $\vert C_{\mathcal{TRDF},G}[A] \vert \leq n^2 + n + 1$. Since checking if a function is a minimal \trdf can be done in polynomial time, we can enumerate all elements in polynomial time. 
\end{pf}    

\begin{corollary}\label{cor:poly_delay_enum_trdf_cobi}
    There is a polynomial-delay algorithm enumerating all minimal \trdfs on cobipartite of order $n$ graphs within time $\Oh^*(\sqrt[3]3^n)$.
\end{corollary} 

\end{toappendix}

\subsection{Interval graphs}

In this subsection, we will consider \crdfs on interval graphs. A graph $G=(V,E)$ is \longversion{called }an \emph{interval graph} if for a each $v\in V$ there is an interval $I_v=[\ell_v,r_v]$ such that $\{u,v\}\in E$ \iffl $I_v \cap I_u\neq \emptyset$. We make use  of \emph{locally definable properties} from~\cite{BliCreOli2021}. 

\begin{figure}[bt]
    \centering    	
	\begin{tikzpicture}[transform shape]
			\tikzset{every node/.style={ fill = white,circle,minimum size=0.3cm}}

            \node[rectangle] at (0,-0.7) {$a_i$};
			\node[rectangle] at (-2,-0.7) {$a_{i-1}$};
			\node[rectangle] at (2,-0.7) {$a_{i+1}$};
			\node[rectangle] at (-1,1.2) {$u_{i-1,1}$};
			\node[rectangle] at (-1,-1.2) {$u_{i-1,2}$};
			\node[rectangle] at (1,1.2) {$u_{i,1}$};
			\node[rectangle] at (1,-1.2) {$u_{i,2}$};

			\node[draw] (a1) at (0,0) {2};
			\node[draw] (a0) at (-2,0) {1};
			\node[draw] (a2) at (2,0) {1};
			\node[draw] (u01) at (-1,0.5) {0};
			\node[draw] (u02) at (-1,-0.5) {0};
			\node[draw] (u11) at (1,0.5) {0};
			\node[draw] (u12) at (1,-0.5) {0};
			\node (dots1) at (2.7,0) {$\dots$};
			\node (dots1) at (-2.7,0) {$\dots$};
   
            \path (a0) edge[-] (u01);	
            \path (a0) edge[-] (u02);
            \path (a1) edge[-] (u01);	
            \path (a1) edge[-] (u02);
            \path (a1) edge[-] (u11);	
            \path (a1) edge[-] (u12);
            \path (a2) edge[-] (u11);	
            \path (a2) edge[-] (u12); 
            
            \path (u01) edge[-] (u02);	
            \path (u11) edge[-] (u12); 
        \end{tikzpicture}

    \caption{Construction of $G_n$ for $i\in \{2i \mid i\in \left[\lfloor\frac{n}{2}\rfloor\right]\}$.}
    \label{fig:nrdp_interval_counterexample}
\end{figure}
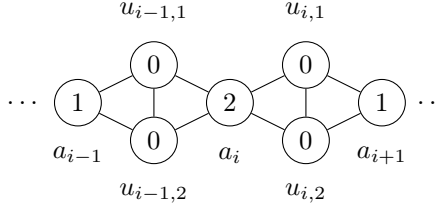
\noindent
Before going into details, we show that we cannot apply \autoref{lem:nrdp_inherit_rdf_bounds} in this case:

\begin{example}
We consider the graph $G_n\coloneqq (V_n,E_n)$ with 
\begin{equation*}
    \begin{split}
        V_n \coloneqq{}& \{a_i\mid i\in [n]\} \cup \{u_{i,j} \mid i \in [n - 1], j\in [2]\},\\
        E_n \coloneqq{}& \{\{a_i,u_{i,j}\}\mid i \in [n-1], j\in [2] \} \cup \{\{a_i,u_{i-1,j}\}\mid i \in [n]\setminus \{1\}, j\in [2] \}.
    \end{split}
\end{equation*} 
$G_n$ is a\shortversion{n}\longversion{ unit} interval graph with the representation $I_{a_i}=[i-0.3, i+0.3]$ and $I_{u_{k,j}}=[k+0.2,k+0.8]$ for $i\in[n], k\in [n-1], j\in[2]$. Furthermore, $G_n$ is planar which can be observed in \autoref{fig:nrdp_interval_counterexample}. 
We define $A\coloneqq\left\{a_{2i} \mid i\in \left[\lfloor\frac{n}{2}\rfloor\right]\right\}$. Hence, $N(A)=\{u_{i,j}\mid i \in [n - 1], j\in [2]\}$. For $f_A\coloneqq 2\chi_{A} + \chi_{V \setminus N[A]}$, $V_1(f_A) \cup V_2(f_A)$ is an independent set. To extend $f_A$ to  minimal \crdf, we must add for all $i \in [n-1]$ exactly one of $u_{i,1}$ and $u_{i,2}$ to $V_1(f_A)$. Therefore, $\vert C_{G_n,\mathcal{CRDF}}[A]\vert\geq\sqrt[3]{2}^n$.
\end{example}

For our enumeration algorithm, we order the vertices by $\ell_v$ (for all $v,u \in V$, $v \leq u$ \iffl $\ell_v \leq \ell_u$). Let $A \subseteq V$. Define $f_A \coloneqq 2\chi_{A} + \chi_{V \setminus N[A]}$. We call a set $X\subseteq N(A)\setminus A$, \emph{$f_A$-connected} \iffl $f_A+\chi_X$ is a minimal \crdf.
\begin{toappendix}

\shortversion{\subsection{$f_A$-connected as Locally Definable Property}}

In this section, we want to prove that $f_A$-connected is a locally definable property. This concept was introduced in \cite{BliCreOli2021}.

Let $(\mathcal{G},\leq)$ be a uniformly linearly ordered set of graphs. For a $G=(V,E)\in \mathcal{G}$, $X\subseteq V$ and $k\in [\vert X\vert]$, $$\mathbb{W}^k(X)\coloneqq\{(x_1,\ldots,x_k)\in X^k\vert \forall i\in [k-1], x\in X\setminus\{x_i,x_{i+1}\}: \neg(x_i\leq x\leq x_{i+1})\}$$ 

A locally definable property on $(\mathcal{G},\leq)$ $\mathcal{P}$ is a uniform vertex set property if there is a $k\in \mathbb{N}\setminus \{0\}$ and three uniform vertex constraints $\phi, \phi^{\min}, \phi^{\max}$ of respective arities $k, k-1,k-1$ such that for any $G=(V,E)\in \mathcal{G}$ and $X\subseteq V$ with $\vert X\vert\geq k$, $X$ has property $\mathcal{P}$ with respect to $G$ \iffl:
\begin{enumerate}
    \item $\forall (x_1,\ldots,x_{k})\in \mathbb{W}^k(X)$, $(x_1,\ldots,x_{k})\in \phi$,
    \item $\forall (x_1,\ldots,x_{k-1})\in \mathbb{W}^{k-1}(X)$:\begin{itemize}
        \item $(\forall x\in X: x_1\leq x) \Rightarrow (x_1,\ldots,x_{k-1})\in \phi^{\min}$,
        \item $(\forall x\in X :x\leq x_{k-1}) \Rightarrow (x_1,\ldots,x_{k-1})\in \phi^{\max}$,
    \end{itemize} 
\end{enumerate}

Now we want to define $\phi, \phi^{\min}, \phi^{\max}$:

Let $s\coloneqq\arg\min \{l_v\mid v\in V_1(f_A) \cup V_2(f_A)\}$ and $t = \arg\max \{r_v\mid v\in V_1(f_A) \cup V_2(f_A)\}$. We define $\phi_{G,A}^{min},\phi_{G,A}^{max}\subseteq  (N(A)\setminus A)^3$ and $\phi_{G,A}\subseteq  (N(A)\setminus A)^4$ as follows:
\begin{enumerate}
    \item $(x,y,z)\in \phi_{G,A}^{min}$ \iffl for all $v \in A$, $P_{G[V\setminus (V_1(f_A) \cup \{x,y,z\})],A}\left( v \right) \nsubseteq \lbrace v\rbrace$ and $s,z$ are in one connected component of $V_1(f_A) \cup A \cup \{s,x,y,z\}$ but not in one connected component of  $V_1(f_A) \cup A \cup \{s,x,z\}$ and not in one connected component of $V_1(f_A) \cup A \cup \{s,y,z\}$.
    \item $(x,y,z)\in \phi_{G,A}^{max}$ \iffl for all $v \in A$, $P_{G[V\setminus (V_1(f_A) \cup \{x,y,z\})],A}\left( v \right) \nsubseteq \lbrace v\rbrace$ and $t,x$ are in one connected component of $V_1(f_A) \cup A \cup \{x,y,z,t\}$ but not in one connected component of $V_1(f_A) \cup A) \cup \{x,z,t\}$ and not in one connected component of $V_1(f_A) \cup V_2(f_A) \cup \{x,y,t\}$.
    \item $(w,x,y,z)\in \phi_{G,A}$ \iffl for all $v \in A$, $P_{G[V\setminus (V_1(f_A) \cup \{w,x,y,z\})],A}\left( v \right) \nsubseteq \lbrace v\rbrace$ and $w,z$ are in one connected component of $V_1(f_A) \cup A \cup \{w,x,y,z\}$ but not in one connected component of $V_1(f_A) \cup A \cup \{w,x,z\}$ and  not in one connected component of $V_1(f_A) \cup A \cup \{w,y,z\}$.
\end{enumerate}

\begin{lemma}
    Let $G=(V,E)$ be an interval graph, $k\in \mathbb{N}\setminus [3]$, $A \subseteq V$ and $X = \{x_1,\ldots,x_k\} \subseteq N(A) \setminus A$ with $x_i<x_j$ \iffl $i<j$ for $i,j\in [k]$. $X$ is $f_A$-connected \iffl $(x_1,x_2,x_3)\in \phi_{G,A}^{min}$, $(x_{k-2},x_{k-1},x_k)\in \phi_{G,A}^{max}$ and for $i\in [k-3]$, $(x_i,x_{i+1},x_{i+2},x_{i+3}) \in \phi_{G,A}$.    
\end{lemma}

\begin{pf}
    Define $g=f_A + \chi_X$.
    Let $X$ be $f_A$-connected. Hence $g$ is a minimal \crdf. Hence, there is an $x\in P_{G[V \setminus V_1(g)],V_2\left(g\right)}\left( v \right) \setminus \{v\}\subseteq P_{G[V \setminus (V_1(f_A)\cup T)],V_2\left(g\right)}\left( v \right) \setminus \{v\}$ for each $T\subseteq X$ and $v\in A = V_2(f_A)=V_2(g)$. This leaves to prove the connectivity property of $\phi_{G,A},\phi_{G,A}^{min},\phi_{G,A}^{max}$. We will only consider~$\phi_{G,A}$. The other ones are analogous. Let $i\in [k-3]$. Assume~$x_i$ and~$x_{i+3}$ are in the same connected component in $V_1(f_A)\cup V_2(f_A) \cup \{x_i,x_{i+2},x_{i+3}\}$. Hence, there is a path $x_i=v_1,\ldots,v_{\alpha}=x_{i+3}$ in $V_1(f_A)\cup V_2(f_A) \cup \{x_i,x_{i+2},x_{i+3}\}$. Obviously, the union of two intersecting intervals is also an interval. By an inductive argument, $[\ell_{x_i},r_{x_{i+3}}] \subseteq \bigcup_{j=1}^{\alpha}I_{v_j}$. Because of minimality of $g$, for all $p,q\in [k]$ with $p\neq q$, $I_{x_{p}}\nsubseteq I_{x_{q}}$. Hence, $I_{x_{i+1}}\subseteq[\ell_{x_i},r_{x_{i+3}}]$, which implies $N[x_{i+1}]\subseteq N[\{v_1,\ldots,v_\alpha\}]$. This contradicts the minimality of~$g$. Thus, $g$ fulfills the properties.
    
    Let $X$ fulfill $(x_1,x_2,x_3)\in \phi_{G,A}^{min}$, $(x_{k-2},x_{k-1},x_k)\in \phi_{G,A}^{max}$ and for $i\in [k-3]$, $(x_i,x_{i+1},x_{i+2},x_{i+3})\in \phi_{G,A}$. By an inductive argument on the connectivity property of $\phi_{G,A}^{min},\phi_{G,A}^{max}$ and $\phi_{G,A}$, $s,t,x_1,\ldots,x_k$ are in one connected component of $V_1(g) \cup V_2(g)$. 
    Therefore, there is a path between $s,t$ in $V_1(g) \cup V_2(g)$ and $[\ell_s,r_t]\subseteq  \bigcup_{v\in V_1(g)\cup V_2(g)}I_v$. Since $s,t$ are minimal respectively maximal, $V_1(g) \cup V_2(g)$ is connected and $g$ a \crdf. No element in $V_1(g)\setminus X$ has a neighbor in~$A$. The elements in $X$ are dominated. Assume there is an $i\in [k-2]\setminus \{1\}$ such that $V_1(g) \cup V_2(g)\setminus \{x_i\}$ is connected. Since $(x_{i-1},x_i,x_{i+1},x_{i+2}) \in \phi_{G,A}$, $V_1(f_A) \cup V_2(f_A) \cup \{x_{i-1},x_{i+1},x_{i+2}\}$ is not connected. Thus, there is a  path between $x_{i-1}$ and $x_{i+1}$ in $V_1(g) \cup V_2(g)\setminus \{x_i\}$, including at least one element in $X\setminus \{x_{i-1},x_i,x_{i+1},x_{i+2}\}$. Hence,    
    $$I_{x_i} \setminus \bigcup_{v \in V_1(f_A) \cup V_2(f_A) \cup \{x_{i-1},x_{i+1},x_{i+2}\}} I_v\subseteq \bigcup_{v \in X \setminus  \{x_{i-1},x_i,x_{i+1},x_{i+2}\}} I_v. $$ 
    This implies that there is a $j\in [i-2]$ such that $I_{x_{i-1}}\subseteq I_{x_j}$. Therefore, either $I_{x_{j+1}}\subseteq I_{x_j}$ or $I_{x_{i-1}}\subseteq I_{x_{j+1}}$. In the first case, $V_1(f_A) \cup V_2(f_A) \cup \{x_{j},x_{j+2},x_{j+3}\}$ is still connected, which contradicts $(x_j,x_{j+1},x_{j+2},x_{j+3})\in \phi_{G,A}$. In the second case, either $j+1=i$ or we continue the argument with $x_{j+1}$. If $j+1=i$, $V_1(f_A) \cup V_2(f_A) \cup \{x_{j},x_{j+2},x_{j+3}\}$ is still connected, which contradicts~$\phi_{G,A}$. For $i =1$, the argument is the same with $x_0\coloneqq s$. For $i\in \{k-1,k\}$, consider the intervals after multiplying with $-1$. 
    This leaves to prove the private neighborhood condition for all $a\in A$. First, there is no $i\in [k]$ with $I_{x_i}\subseteq I_{a}$, as otherwise $ V_1(f_A) \cup V_2(f_A) \cup \{x_{i-1},x_{i+1},x_{i+2}\}$ is still connected, which contradicts~$\phi_{G,A}$. 
    Therefore, either $\ell_a<\ell_{x_1}$, $r_{x_k}<r_a$, or there is an $i\in [k-1]$ such that $I_a\subseteq [\ell_{x_i},r_{x_{i+1}}]$.
    We assume the last case. The other ones are analogous. Hence, there exists a $v\in P_{G[V\setminus (V_1(f) \cup \{x_{i-1},x_{i},x_{i+1},x_{i+2}\})],V_2\left(f\right)}\left( a\right)\setminus \{a\}$. As in the  connectivity property case, there are no elements $p,q \in [k]$ such that $I_{x_p}\subseteq I_{x_q}$. Therefore, $v\notin X$. Thus, for all $a\in A$, $ P_{G',V_2\left(f\right)}\left( v \right) \nsubseteq \lbrace v\rbrace$.
\end{pf}

As membership in $\phi_{G,A}^{min},\phi_{G,A}^{max}$ and $\phi_{G,A}$ can be decided in polynomial time, $f_A$-connectivity is a locally definable property on interval graphs.    
\end{toappendix}
\shortversion{As $f_A$-connectivity is a locally definable property on interval graphs $(\star)$
, w}\longversion{W}ith \cite[Theorem 3.2]{BliCreOli2021}, we can deduce:  

\begin{corollary}
    Let $G=(V,E)$ be an interval graph and $A\subseteq V$. The set of all $f_A$-connected sets can be enumerated with polynomial delay.  
\end{corollary}

Since there is a bijection between the $f_A$-connected sets on $G$ and $C_{G,\mathcal{CRDF}}[A]$ which can be calculated in polynomial time, we obtain:

\begin{theorem}
    There is a polynomial-delay algorithm for enumerating all minimal \crdfs on interval graphs.
\end{theorem}

\begin{remark}\label{rem:nrdp_with_different_approach}
    The approach described in \cite{BliCreOli2021} use of an existing polynomial delay algorithm for enumerating paths on directed graphs. Therefore, in this subsection we consider the combination of our (extension) approach together with this one. This is another strength of the \nrdp approach. By considering the vertices of value~1 after fixing the vertices of value~2, we can combine different approaches. Even if we do not consider further such cases it would be interesting to see more such combinations (for example with the supergraph method \cite{KobKMMO2025,KurWas2024}).  
\end{remark}

\section{Downsides of the Approach}\label{sec:downsides_nrdp}

Is this section, we show that using this method naively is not helpful every time. To this end, we define  the following problem for any \nrdp $\mathcal{P}$:

\centerline{\fbox{\begin{minipage}{.99\textwidth}
\textbf{Problem name: }$\mathcal{P}$-\textsc{Non-Emptiness}\\
\textbf{Given: } Graph $G=(V,E)$ and $A \subseteq V$.\\
\textbf{Question: } Is $C_{\mathcal{P},G}[A]\neq \emptyset$?\end{minipage}
}}

\begin{theorem}\label{thm:crdf_paranp_maxdeg4}
    $\mathcal{CRDF}$-\textsc{Non-Emptiness} is \NP-complete even on bipartite 2-degen\-er\-ate graphs of maximum degree~4. \longversion{$\vert V_1(f)\vert$-\ECRD is \paraNP-hard even on bipartite graphs of maximum degree~4. }
\end{theorem}
\begin{pf}
    For the hardness we use \textsc{Monotone 3-Sat-(2,2)}. This is a variation of \textsc{3-Sat}, where each variable appears twice positively and twice negatively. \cite{DarDoc2021} provides an \NP-completeness proof for this problem. So, let $X \coloneqq \{x_1,\ldots,x_n\}$ be the set of variables and $C \coloneqq \{c_1,\ldots,c_m\}$ the set of clauses. To simplify the notation, denote $X'=\{x_i,\overline{x_i} \mid i\in [n]\}$. We can assume that there is no pair $(i,j)\in [n]\times [m]$  such that $\{x_i,\overline{x}_i\} \subseteq c_j$. Otherwise, we can\longversion{ just} delete $c_j$ from~$C$. For each $j\in [m]$, let $c_j=\{l_{j,1}, l_{j,2}, l_{j,3}\}\in C$ with $l_{j,1}, l_{j,2}, l_{j,3} \in X'$.  

    Define $G=(V,E)$ with 
    \begin{equation*}
        \begin{split}
            V \coloneqq {}& \{v_i,\overline{v_i},w_i\mid i\in [n]\} \cup \{ p_{j}\mid j\in [m]\} \cup \{u_i,u'_i\mid i\in [n-1]\},\\
            E \coloneqq {}& \{\{v_i,w_i\},\{\overline{v_i}, w_i\}\mid i\in [n]\}  \cup \{\{u_{i},w_{i}\}, \{u_{i},w_{i+1}\},\{u_{i},u'_{i}\} \mid i \in [n-1] \} \cup{}\\
            &\{ \{v_i,p_{j}\} \mid j\in [m], i\in [n],\exists k\in [3]:\: x_i=l_{j,k}\}\cup{}\\
            &\{\{\overline{v_i},p_{j}\} \mid j\in [m], i\in [n],\exists k\in [3]:\: \overline{x_i}=l_{j,k}\}.
        \end{split}
    \end{equation*}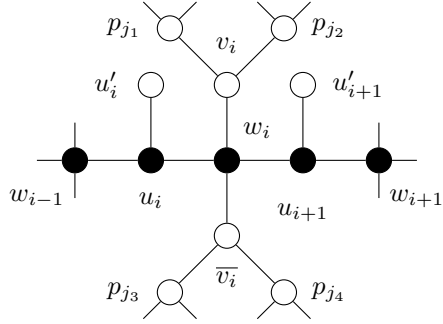
\begin{figure}[t]
    \centering
	\begin{tikzpicture}[transform shape,fill lower half/.style={path picture={\fill[#1] (path picture bounding box.south west) rectangle (path picture bounding box.east);}}]
			\tikzset{every node/.style={ fill = white,circle,minimum size=0.3cm}}
			\node  (labelwi) at (0.4,0.4) {$w_{i}$};
			\node  (labelwi-1) at (-2.5,-0.5) {$w_{i-1}$};
			\node  (labelwi+1) at (2.5,-0.5) {$w_{i+1}$};
			\node[draw,label={left:$p_{j_1}$}] (p1) at (-0.75,1.75) {};
			\node[draw,label={right:$p_{j_2}$}] (p2) at (0.75,1.75) {};
			\node[draw,label={left:$p_{j_3}$}] (p3) at (-0.75,-1.75) {};
			\node[draw,label={right:$p_{j_4}$}] (p4) at (0.75,-1.75) {};
			\node[draw,label={above:$v_i$}] (vi) at (0,1) {};
			\node[draw,label={below:$\overline{v_i}$}] (v'i) at (0,-1) {};
			\node[draw,fill=black] (wi) at (0,0) {};
			\node[draw,fill=black] (wi-1) at (-2,0) {};
			\node[draw,fill=black] (wi+1) at (2,0) {};
			\node[draw,fill=black,label={below:$u_i$}] (ui) at (-1,0) {};
			\node[draw,fill=black,label={below:$u_{i+1}$}] (ui+1) at (1,0) {};
			\node[draw,label={left:$u'_i$}] (u'i) at (-1,1) {};
			\node[draw,label={right:$u'_{i+1}$}] (u'i+1) at (1,1) {};
            
    		\path (2,0.5) edge[-] (wi+1);
    		\path (-2,0.5) edge[-] (wi-1);
    		\path (2,-0.5) edge[-] (wi+1);
    		\path (-2,-0.5) edge[-] (wi-1);
    		\path (2.5,0) edge[-] (wi+1);
    		\path (-2.5,0) edge[-] (wi-1);
    		\path (wi) edge[-] (ui+1);
    		\path (wi) edge[-] (ui);
    		\path (wi+1) edge[-] (ui+1);
    		\path (wi-1) edge[-] (ui);    		
            \path (u'i+1) edge[-] (ui+1);
    		\path (u'i) edge[-] (ui);
    		\path (wi) edge[-] (vi);
    		\path (wi) edge[-] (v'i);
    		\path (p1) edge[-] (vi);
    		\path (p2) edge[-] (vi);
    		\path (p3) edge[-] (v'i);
    		\path (p4) edge[-] (v'i);
    		\path (p1) edge[-] (-1.1,2.1);
    		\path (p1) edge[-] (-0.4,2.1);
    		\path (p2) edge[-] (1.1,2.1);
    		\path (p2) edge[-] (0.4,2.1);
    		\path (p3) edge[-] (-1.1,-2.1);
    		\path (p3) edge[-] (-0.4,-2.1);
    		\path (p4) edge[-] (1.1,-2.1);
    		\path (p4) edge[-] (0.4,-2.1);
        \end{tikzpicture}

    \caption{Construction of \autoref{thm:crdf_paranp_maxdeg4} where the black vertices are in $A$.}
    \label{fig:crdf_paranp_maxdeg4}
\end{figure}
    This graph has the partition classes $\{v_i,\overline{v}_i\mid i\in [n]\} \cup \{u_i\mid i\in [n-1]\}$ and $\{w_i\mid i\in [n]\} \cup \{u'_i\mid i\in [n-1]\} \cup \{p_{j}\mid j\in [m]\}$. 
    Thus, $G$ is a bipartite graph. Now we will show that $G$ is 2-degenerate. At first, for all $i\in [n-1]$, $\deg(u'_{i})=1$. After deleting these vertices from $G$, $u_i$ has degree~2 for all $i\in [n-1]$ ($u_i$ has degree~3 in $G$). Let $G'$ be the graph after deleting also these vertices. Now let $i\in [n]$. While $w_{i}$ has degree at most~4 in~$G$, the degree in~$G'$ is~2. After deleting $\{w_r \mid r\in [n]\}$, $v_i,\overline{v_i}$ have degree~2. In~$G$, $v_i,\overline{v_i}$ have degree~3. Therefore, we only need consider~$p_j$ for $j\in [m]$. But $G[\{p_j\mid j\in [m]\}]$ has no edges. Hence, $G$ is a 2-degenerate graph with maximum degree~4. 
    
    Let $A\coloneqq \{w_i\mid i\in [n]\} \cup \{ u_i \mid i\in [n-1]\}$. 
    The gadget is also depicted in \autoref{fig:crdf_paranp_maxdeg4}.
    \begin{claim}
        There is an assignment $\phi$ satisfying $C$ \iffl $C_{\mathcal{CRDF},G}[A]\neq \emptyset$.
    \end{claim}
    
    \begin{pfclaim}
        Let $\phi$ be a satisfying assignment. Let $I=[n]$. If there is an $i\in I$, such that $\phi-(1-\phi(x_i))\chi_{\{ x_i \}}$ is still satisfying $C$, delete $i$ from $I$. \shortversion{Since}\longversion{By the fact that} there is no pair $(i,j)\in [n]\times [m]$ with $\{x_i,\overline{x}_i\} \subseteq C_j$, $\phi$ restricted to $I$\longversion{ already} satisfies $C$. Define
        $h\coloneqq 2\cdot \chi_{A}+ \chi_{\{v_i\mid  i\in I, \phi(x_i)=1\} \cup \{\overline{v_i}\mid  i\in I, \phi(x_i)=0\}} + \chi_{\{p_1,\ldots,p_m\}}\,.$
        Clearly, $h$ is a \rdf, as $V\setminus N(A)=\{p_1,\ldots,p_m\}$. 
        As $\phi$ satisfies $C$, \longversion{$N(p_{j})\cap V_1(h)\neq \emptyset$ for all $j\in [m]$. Thus,} for each $j\in [m]$ there is an $i\in I$ such that $w_i$ is connected to $p_j$ in $V_1(h) \cup V_2(h)$. \longversion{For $k,\ell\in [n]$ with $k<\ell$, $w_k$ and $w_{\ell}$ are connected by $w_k,u_k,w_{k+1}, \ldots, w_{\ell}$.}\shortversion{Furthermore, $A$ is connected.} Hence, $V_1(h) \cup V_2(h)$ is connected and $h$ a \crdf.
        Next, we want to prove that each element in $V_2(h)=\{w_1,u_1,w_2,\ldots,u_{n-1},w_n\}$ has a private neighbor. Since $\phi$ is assignment, $\emptyset\neq \{v_i, \overline{v}_i\}\cap V_0(h) = N(w_i)\cap V_0(h)$ for all $i \in [n]$. Further, $u_i'$ is the private neighbor of $u_i$ for $i\in [n-1]$. This implies Property \ref{con_crdf_private} of \shortversion{\autoref{thm:minimal_tcrdf}}\longversion{\autoref{lem:minimal_crdf}}. For $j\in [m]$, $N(p_j)\cap V_2(h)= \emptyset$.  This leaves only to show that  $G[(V_1(h) \cup V_2(h))\setminus \{t\}]$ is not connected for any $t\in \{v_i,\overline{v}_i\mid i\in I\} \cap V_1(h)$. Let $i\in I$ and \swlog $h(v_i)=1$ (for $h(\overline{v}_i)=1$ works analogously). By definition of $I$ there is a $j\in [m]$ such that $C_j$ is only satisfied by $x_i$. Hence, $p_j$ has no neighbor in $V_1(h-\chi_{\{v_i\}}) \cup V_2(h-\chi_{\{v_i\}})$. Therefore, $h$ is a minimal \crdf.
    
        Conversely, assume there is a minimal \crdf $g\in \{0,1,2\}^V$ with $f\leq g$. Define $\phi \coloneqq \chi_{\{x_i\mid i\in [n], g(v_i) \neq 0 \}}$.
        Let $i\in [n]$. Since $w_i\in V_2(f)\subseteq V_2(g)$, it needs a private neighbor and $\{v_i,\overline{v}_i\}\cap V_0(g)\neq \emptyset$.
        Let $j\in [m]$. \Wlog, we consider~$C_j$ includes only negative literals. Since $g$ is a \crdf, there is an $i\in [n]$ with $\overline{v}_i\in N(p_j)\cap (V_1(g) \cup V_2(g))$. Hence,\longversion{ $g(v_i)=0$, $\phi(x_i)=0$ and} $\phi$ satisfies~$C_j$.
    \end{pfclaim}
    
    Since the graph has order $3n + m$ and $6n$ edges, $G$ and $f$ can be constructed in polynomial time. 
\end{pf}

\noindent
There are simular results for \trdfs: 

\begin{theorem}\label{thm:trdf_paranp_maxdeg3}\shortversion{$(\star)$}
    $\mathcal{TRDF}$-\textsc{Non-Emptiness} is \NP-complete even on bipartite graphs of maximum degree~3. 
\end{theorem}
 \begin{toappendix}
\begin{pf}\shortversion{(of \autoref{thm:trdf_paranp_maxdeg3})}
    For the hardness we use \textsc{Monotone 3-Sat-(2,2)}. It is known by \cite{DarDoc2021}, that this restricted \textsc{Sat} version is \NP-complete. So, let $X \coloneqq \{x_1,\ldots,x_n\}$ be the set of variables and $C \coloneqq \{C_1,\ldots,C_m\}$ the set of clauses. To simplify the notation, denote $X'=\{x_i,\overline{x_i} \mid i\in [n]\}$. For each $j\in [m]$, let $C_j=l_{j,1}\vee l_{j,2} \vee l_{j,3}\in C$ with $l_{j,1}, l_{j,2}, l_{j,3} \in X'$.
    
\begin{figure}[bt]
    	\centering
	\begin{tikzpicture}[transform shape]
		      \tikzset{every node/.style={ fill = white,circle,minimum size=0.3cm}}
			\node[draw,label={below:$v_i$}] (vi) at (-1,0) {0};
			\node[draw,label={below:$\overline{v_i}$}] (v'i) at (1,0) {0};
			\node[draw,label={below:$w_i$}] (wi) at (0,0) {2};
   		  \node[draw,label={below:$p_{j_1}$}] (p1) at (-1.75,0.75) {0};
   		  \node[draw,label={below:$p_{j_2}$}] (p2) at (-1.75,-0.75) {0};
   		  \node[draw,label={below:$p_{j_3}$}] (p3) at (1.75,0.75) {0};
   		  \node[draw,label={below:$p_{j_4}$}] (p4) at (1.75,-0.75) {0};
   		   
    		\path (wi) edge[-] (vi);
    		\path (wi) edge[-] (v'i);
    		\path (p1) edge[-] (vi);
    		\path (p2) edge[-] (vi);
    		\path (p3) edge[-] (v'i);
    		\path (p4) edge[-] (v'i);

    		\path (p1) edge[-] (-2.15,0.4);
    		\path (p1) edge[-] (-2.15,1.1);
            
    		\path (p2) edge[-] (-2.15,-0.4);
    		\path (p2) edge[-] (-2.15,-1.1);
            
    		\path (p3) edge[-] (2.15,0.4);
    		\path (p3) edge[-] (2.15,1.1);
            
    		\path (p4) edge[-] (2.15,-0.4);
    		\path (p4) edge[-] (2.15,-1.1);
        \end{tikzpicture}
    \caption{Construction for \autoref{thm:trdf_paranp_maxdeg3}}
 \label{fig:trdf_paranp_maxdeg3}
\end{figure} 
    Define $G=(V,E)$ with 
    \begin{equation*}
        \begin{split}
            V \coloneqq{} & \{v_i,\overline{v_i},w_i\mid i\in [n]\} \cup \{ p_j\mid j\in [m]\}\\
            E \coloneqq{}& \{\{v_i,w_i\},\{\overline{v_i}, w_i\}\mid i\in [n]\}  \cup \\
            &\{\{v_i,p_j\}\mid j\in [m], k\in [3], i\in [n], x_i=l_{j,k}\}\cup{}\\
            &\{\{\overline{v_i},p_j\} \mid j\in [m], k\in [3], i\in [n], \overline{x_i}=l_{j,k}\}.
        \end{split}
    \end{equation*}
    This graph has the partition classes $\{v_i,\overline{v}_i\mid i\in [n]\}$ and $\{w_i\mid i\in [n]\} \cup \{p_j \mid j\in [m]\}$. Because of the special \textsc{Sat} variant we consider, for all $i\in[n] $ and $j\in [m]$,
    $$\deg(w_i)=2<3=\deg(v_i)=\deg(\overline{v}_i)=\deg(p_j).$$
    Thus $G$ is a bipartite graph with degree at most~$3$.
    Let $A\coloneqq \{w_i\mid i\in [n]\}$ and $f \coloneqq 2 \cdot \chi_{A} \in \{0,1,2\}^V$. 
    The gadgets and the values of $f$ are also depicted in \autoref{fig:trdf_paranp_maxdeg3}.
    \begin{claim}
        There is an assignment $\phi$ satisfying $C$ \iffl $C_{\mathcal{TRDF},G}[A]\neq \emptyset$ \iffl there is a minimal \trdf $h$ on $G$ with $f \leq h$.
    \end{claim}
    \begin{pfclaim}
        By \autoref{lem:NRDP_G_eq_extension}, $C_{\mathcal{TRDF},G}[A]\neq \emptyset$ implies the existence of a minimal \trdf $h$ on $G$ with $f \leq h$.
    
        Let $\phi$ be a satisfying assignment. Then define $$h\coloneqq f+ \chi_{\{v_i\mid  i\in [n], \phi(x_i)=1\} \cup \{v_i\mid  i\in [n], \phi(x_i)=0\}} + \chi_{\{ p_j\mid j\in [m]\}}\,.$$ Clearly $V_0(h) \subseteq \{v_i, \overline{v}_i \mid i\in [n]\}\subseteq N(\{ w_i \mid  i\in [n]\})\subseteq N(V_2(h))$. Therefore, $h$ is a \rdf. 
        Since $\phi$ is an assignment, for each $i\in [n]$, there is $u_i\in N(w_i)\cap V_1(h)$. Further $\phi$ satisfies $C$. Hence, for all $j\in [m]$, $N(p_j)\cap V_1(h)\neq \emptyset$. Thus, $h$ is a \trdf.
        Since $V_2(h))=\{w_i\mid i\in [n]\}$ and $\phi$ is assignment, $\emptyset\neq \{v_i, \overline{v}_i\}\cap V_0(h) = N(w_i)\cap V_0(h)$. This implies Property~\ref{con_trdf_private} of \shortversion{\autoref{thm:minimal_tcrdf}}\longversion{\autoref{lem:minimal_trdf}}. For $j\in [m]$, $N(p_j)\cap V_2(f)= \emptyset.$ This leaves only to show that  $G[(V_1(h) \cup V_2(h))\setminus \{u\}]$ has an isolated vertex for $u\in \{v_i,\overline{v}_i\mid i\in [n]\} \cap V_1(h)$. Let $i\in[n] $ and \swlog $f(v_i)=1$. As $\phi$ is an assignment, $f(\overline{v}_i)=0$. Because $\overline{v}_i$ is besides $v_i$ the only neighbor of $w_i$, $w_i$ would be isolated of $G[(V_1(h) \cup V_2(h))\setminus \{v_i\}]$. By \shortversion{\autoref{thm:minimal_tcrdf}}\longversion{\autoref{lem:minimal_trdf}}, $h$ is a minimal \trdf. Furthermore, $f\leq h\in C_{\mathcal{TRDF},G}[A]$.
    
        Conversely, assume there is a $g\in \{0,1,2\}^V$ be a minimal \trdf with $f\leq g$. Define $\phi \coloneqq \chi_{\{x_i\mid i\in [i], g(v_i)\neq 0 \}}$.
        Let $i\in [n]$. Since $w_i\in V_2(f)\subseteq V_2(g)$, it needs a private neighbor and $\{v_i,\overline{v}_i\}\cap V_0(g)\neq \emptyset$. Further, $N(w_i)\cap (V_1(g)\cup V_2(g))\neq \emptyset$, as otherwise $g$ would not be a \trdf.   

        Let $j\in [m]$. \Wlog, we consider $C_j$ includes only negative literals. Assume $g(p_j)=0$. Then there exists an $i\in [n]$ with $\overline{v}_i\in N(p_j)\cap V_2(g)$. Hence, $g(v_i)=0$ and $\phi(x_i)=0$ and~$\phi$ satisfies~$C_j$. This leaves to consider $g(p_j)\neq 0$. As $g$ is a \trdf, there is a $\overline{v}_i\in N(p_j)\cap (V_1(g) \cap V_2(g))$ and analogously to $g(p_j)=0$, $\phi$ satisfies~$C_j$.
    \end{pfclaim}
    
    Since the graph has order $3n + m$ and $6n$ edges, $G$ and $f$ can be constructed in polynomial time. 
\end{pf}
\end{toappendix}

\noindent
We have even some hardness results for split graphs. For the proof we use the following lemma:

\begin{lemma}\label{lem:bij_crdf_trdf_split}\shortversion{$(\star)$}
    Let $G=(V,E)$ be a connected split graph without universal vertices. Then $f\in \{0,1,2\}^V$ is a \trdf on $G$ \iffl $f$ is a \crdf. 
\end{lemma}
\begin{toappendix}
\begin{pf}\shortversion{(of \autoref{lem:bij_crdf_trdf_split})}
    Let $G=(V,E)$ be split graph with the maximal clique $C$ and independent set $I$ as well as $f\in \{0,1,2\}^V$. Define $V'\coloneqq V_1(f) \cup V_2(f)$ and $G'\coloneqq G[V']$. Let $f$ be a \trdf.  Let $v,u\in V'$ with $v\neq u$. Let $v,u\in I$. 
    Since $f$ is a \trdf there exists an $x\in N(v)\cap V'\subseteq C \cap V'$. With the same argument, there is a $y\in N(u) \cap V' \subseteq C \cap V'$. If $x=y$, $vxu$ forms a path on $G'$. Otherwise, $vxyu$ forms a path.  
    Analogously, there is a path from $v$ to $u$ if $\{v,u\}\cap C \neq \emptyset$. Therefore, $G'$ is connected. 
    
    Now we consider $f$ is a connected \rdf on $G$. Since there is no universal vertex, there is no vertex $v\in V$ such that $\{v\}$ is a dominating set. As $V'$ is a dominating set, $\vert V'\vert\geq 2$. Hence, each $v\in V'$ has a neighbor in $V'$ and $f$ is total.
\end{pf}
\end{toappendix}

\begin{theorem}\label{thm:nrdp_crdf_split}\shortversion{$(\star)$}
    \textsc{Enum} $C_{\mathcal{CRDF}}$ and \textsc{Enum} $C_{\mathcal{TRDF}}$ are at least as hard as \textsc{Enum Tr} even on split graphs.
\end{theorem}

\shortversion{To clarify, why these hardness result are limitations to our approach, we present the construction of the proof. The complete prove can be found in \todo{Arxiv}.
\begin{pf}
Let $H=(X,S)$ be the hypergraph with $X \coloneqq \{x_1,\ldots,x_n\}$ and $S\coloneqq\{s_1,\ldots,s_m\}$. We can assume that there is no  $i\in [n]$ such that $x_i\in s_j$ for all $j\in [m]$. Otherwise, $Tr(G)=\{\{x_i\}\} \cup Tr((X\setminus \{x_i\},S ))$. Define $G=(V,E)$ with 
    \begin{equation*}
        \begin{split}
            V \coloneqq{}& \{a,b\} \cup \{u_i\mid i\in [n]\} \cup \{w_j\mid j\in [m] \}\\
            E \coloneqq{}& \binom{\{a,b\} \cup  \{u_i\mid i\in [n]\}}{2} \cup \{\{u_i,w_j\}\mid i\in [n], j\in [m], x_i\in s_j\}.
        \end{split}
    \end{equation*}
    \begin{claim}
        $F: C_{\mathcal{CRDF},G}[\{a\}] \to Tr(H),\, g\mapsto \{x_i\mid i\in [n], g(u_i)=1\}$ is a bijection.
    \end{claim}
\end{pf}}
    
\begin{toappendix}
\begin{pf}\shortversion{(of \autoref{thm:nrdp_crdf_split})}
    Let $H=(X,S)$ be the hypergraph with $X \coloneqq \{x_1,\ldots,x_n\}$ and $S\coloneqq\{s_1,\ldots,s_m\}$. We can assume that there is no  $i\in [n]$ such that $x_i\in s_j$ for all $j\in [m]$. Otherwise, $Tr(H)=\{\{x_i\}\} \cup Tr((X\setminus \{x_i\},S ))$. Define $G=(V,E)$ with 
    \begin{equation*}
        \begin{split}
            V \coloneqq{}& \{a,b\} \cup \{u_i\mid i\in [n]\} \cup \{w_j\mid j\in [m] \}\\
            E \coloneqq{}& \binom{\{a,b\} \cup  \{u_i\mid i\in [n]\}}{2} \cup \{\{u_i,w_j\}\mid i\in [n], j\in [m], x_i\in s_j\}.
        \end{split}
    \end{equation*}
    \begin{claim}
        $F: C_{\mathcal{CRDF},G}[\{a\}] \to Tr(H),\, g\mapsto \{x_i\mid i\in [n], g(u_i)=1\}$ is a bijection.
    \end{claim}
    \begin{pfclaim}    
        Let $g \in C_{\mathcal{CRDF},G}[\{a\}]$. By \autoref{lem:NRDP_G_eq_extension}, $2\cdot\chi_{\{a\}} + \chi_{\{w_j\mid j\in [m]\}} \leq g$. Define $Z\coloneqq F(g)$. Since $V_1(g)\cup \{a\}$ needs to be connected and $\{w_1,\ldots,w_j\} \subseteq V_1(g)$, for all $j\in [m]$, $Z\cap s_j\neq \emptyset$. Hence $Z$ is a \hs. Assume $Z$ is not minimal. Then there is an $i \in [n]$ such that for all $j\in [m]$, $s_j \cap (Z\setminus \{x_i\})\neq \emptyset$. Hence, for $g_i\coloneqq g-\chi_{\{u_i\}}$, $V_1(g_i)\cup V_2(g_i)$ is connected. Since~$u_i$ is dominated by~$a$, $g_i$ is a \crdf. This contradicts the minimality of~$g$. Thus, $Z$ is minimal. 

        Conversely, let $D\subseteq X$ be a minimal \hs. Define $g\coloneqq 2\cdot\chi_{\{a\}} + \chi_{\{u_i \mid x_i\in D \}} + \chi_{\{w_j\mid j\in [m]\}}$. Clearly, $D=F(g)$. Since~$D$ is a dominating set, for all $j\in [m]$ there is an $i \in [n]$ with $x_i\in s_j$ and $u_i\in N(w_j)\cap V_1(g)$. As $\{a,b\} \cup  \{u_i\mid i\in [n]\}$ is a clique, $V_1(g) \cup V_2(g)$ is connected. Furthermore, $V_0(g) \subseteq \{a,b\} \cup  \{u_i\mid i\in [n]\} \subseteq N[a]$. Hence, $g$ is a \crdf. $b$ is a private neighbor of~$a$. Assume there is an $i\in [n]$ such that $g_i\coloneqq g - \chi_{\{u_i\}}$ is a \crdf. As $D$ is minimal, there exists a $j\in [m]$ with $s_j \cap D =  \{x_i\}$. Thus, $w_j$ would be isolated in $G[V_1(g_i) \cup V_1(g_i)]$. Furthermore, $\emptyset \neq  \{w_j\mid j \in [m]\}\cap N[a] = \{w_j\mid j \in [m]\}\cap N[V_2(g)]$. Therefore, $g$ is minimal. This implies $F$ is surjective. 
        
        Let $g,h\in C_{\mathcal{CRDF},G}[\{a\}]$ be minimal \crdfs with $D\coloneqq F(g)=F(h)$. Therefore, $V_1(g)\cap \{u_i\mid i\in [n]\}=V_1(h)\cap \{u_i\mid i\in [n]\}$.  Since $V_2(g)= \{a\}=V_2(h)$, $\{w_j\mid j\in [m]\}\subseteq V_1(g) \cap V_1(h)$. This implies $g$ and $h$ can only differ in $b$. If they differ in just one vertex, $h\leq g$ or $g \leq h$. This would contradict the minimality of $g$ or $h$ and implies $g=h$. Hence, $F$ is injective and therefore bijective. 
    \end{pfclaim}

    Since $F$ is bijection for which $F(g)$ can be calculated in polynomial time for all $g\in C_{\mathcal{CRDF},G}[\{a\}]$, we could enumerate all minimal \hs of $H$  in output polynomial time if we could enumerate all functions in $C_{\mathcal{CRDF},G}[\{a\}]$.

    It should be mentioned that $G$ has no universal vertex. Hence, each \crdf is a \trdf and the other way around. 
\end{pf}
\end{toappendix}

\longversion{\begin{corollary}
    $\vert V_2(f)\vert$-\ETRD,$\vert V_2(f)\vert$-\linebreak[4]\ECRD\ are \paraNP-hard on split graphs.
\end{corollary}}

\begin{remark}\label{rem:nrdp_begin_approach}
    The hardness result in this section imply that this approach is not helpful in every case. For the proof of \autoref{thm:nrdp_polyenumeration}, we need a output-polynomial enumeration algorithm for $C_{\mathcal{P}}[A]$. This is not possible, if it is \NP-complete to decide if there is element in $C_{\mathcal{P}}[A]$. 
    
    The algorithm of \autoref{thm:nrdp_polyenumeration} does not consider $C_{\mathcal{P},G}[B]$ for $B\subseteq A$. The supergraph method, for example, uses already existing solutions to compute new once. This implies the question, whether we could use different approaches (besides extension) for results like \autoref{thm:nrdp_polyenumeration}.

    By \autoref{thm:nrdp_crdf_split}, we know, that even the idea of using $C_{\mathcal{CRDF},G}[B]$ for\linebreak[4] $C_{\mathcal{CRDF},G}[A]$, when $B\subseteq A$, is not helpful in every case. In this proof $A=\{a\}$. So, the only proper subset is  $\emptyset$. By the definition,  $C_{\mathcal{CRDF},G}[\emptyset]=\{\chi_V\}$. Hence, using proper subsets is not helpful in every case.
\end{remark}

\section{Conclusion}

We introduced the notion of \nrdp and used it to obtain some enumeration results for different Roman domination variations like \mrdf, \trdf, \crdf.  Furthermore, we proved that our approach is not always helpful unless $\Ptime=\NP$. In these cases, it does not prove that the enumeration is hard. Thus, it would be interesting to prove some results for enumerating minimal \crdfs/\trdfs on general graphs. 
There could be even more graph classes with good enumeration results.
Here, it could be interesting to consider to use different approaches/methods as mentioned in \autoref{rem:nrdp_with_different_approach} and \autoref{rem:nrdp_begin_approach}.
If we have hardness results we could also use the approach to obtain \FPT delay algorithms.
Improving the input-sensitive bounds for enumerating minimal \rdfs could strengthen results implied by \autoref{lem:nrdp_inherit_rdf_bounds}. 

Furthermore, not all Roman domination variations are \nrdp. For example, unique response Roman domination (see \cite{RubSla2007}) or Italian domination (see \cite{CheHHM2016,HenKlo2017}) fail to be. Finding more \nrdps, or even a generalization of \nrdp which also handles some of these examples could be a future research direction. 
Another question is if there are results beyond enumeration which could use \nrdp.
\shortversion{\newpage}

\bibliographystyle{splncs04}
\bibliography{ab,hen}
\end{document}